\documentclass[sn-mathphys-num]{sn-jnl}
\usepackage{graphicx}%
\usepackage{multirow}%
\usepackage{amsmath,amssymb,amsfonts}%
\usepackage{amsthm}%
\usepackage{mathrsfs}%
\usepackage[title]{appendix}%
\usepackage{xcolor}%
\usepackage{textcomp}%
\usepackage{manyfoot}%
\usepackage{booktabs}%
\usepackage{algorithm}%
\usepackage{algorithmicx}%
\usepackage{algpseudocode}%
\usepackage{listings}%

\usepackage{dblfloatfix}

\usepackage[caption=false]{subfig}
\usepackage{dcolumn}
\usepackage{bm}

\DeclareMathOperator\erf{erf}

\DeclareMathOperator\E{E}

\usepackage{hyperref} 

\usepackage{siunitx}
\usepackage{orcidlink}

\usepackage{listings}
\usepackage{color}

\hypersetup
{
	pdfauthor={Weida Liao},
	pdftitle={Axisymmetric thermoviscous and thermal expansion flows for microfluidics}
}

\begin{document}

\title[Axisymmetric thermoviscous and thermal expansion flows for microfluidics]{Axisymmetric thermoviscous and thermal expansion flows for microfluidics}

\author[1]{\fnm{Weida} \sur{Liao}\,\orcidlink{0000-0002-0000-228X}}

\author*[1]{\fnm{Eric} \sur{Lauga}\,\orcidlink{0000-0002-8916-2545}}\email{e.lauga@damtp.cam.ac.uk}

\affil[1]{\orgdiv{Department of Applied Mathematics and Theoretical Physics}, \orgname{University of Cambridge}, \orgaddress{\street{Wilberforce Road}, \city{Cambridge}, \postcode{CB3 0WA},  \country{UK}}}

\abstract{		
	Recent microfluidic experiments have explored the precise positioning of micron-sized particles in liquid environments via laser-induced thermoviscous flow. From micro-robotics to biology at the subcellular scale, this versatile technique has found a broad range of applications. Through the interplay between thermal expansion and thermal viscosity changes, the repeated scanning of the laser along a scan path results in fluid flow and hence net transport of particles, without physical channels. Building on previous work focusing on two-dimensional microfluidic settings, we present an analytical, theoretical model for the thermoviscous and thermal expansion flows and net transport induced by a translating heat spot in three-dimensional, unconfined fluid. We first numerically solve for the temperature field due to a translating heat source in the experimentally relevant limit. Then, in our flow model, the small, localised temperature increase causes local changes in the mass density, shear viscosity and bulk viscosity of the fluid. We derive analytically the instantaneous flow generated during one scan and compute the net transport of passive tracers due to a full scan, up to quadratic order in the thermal expansion and thermal shear viscosity coefficients. We further show that the flow and transport are independent of bulk viscosity. In the far field, while the leading-order instantaneous flow is typically a three-dimensional source or sink, the leading-order average velocity of tracers is instead a source dipole, whose strength depends on the relative magnitudes of the thermal expansion and thermal shear viscosity coefficients. Our quantitative results reveal the potential for future three-dimensional net transport and manipulation of particles at the microscale.
}


\maketitle

\section{Introduction}

The controlled, precise manipulation of microparticles and biological cells in a liquid environment underpins applications across a wide range of disciplines and industries, from drug delivery and disease diagnostics, to the construction of miniaturised devices and the study of single cells and molecules~\cite{cheng2021active,nilsson2009review,zhang2019robotic}. 
Microparticle manipulation has been accomplished using a variety of physical mechanisms to achieve trapping, particle movement or assembly of structures~\cite{zhang2019robotic}. 
Optical tweezers trap particles with optical forces associated with spatial gradients in light intensity, created by focused light~\cite{moffitt2008recent,favre2019optical,palima2013gearing}, while magnetic tweezers apply force to magnetic beads via a magnetic field~\cite{de2012recent}, and acoustic tweezers use the acoustic radiation force induced by standing waves~\cite{baudoin2020acoustic}.
For optothermal micromanipulation, thermal gradients caused by laser heating result in thermophoresis, the migration of objects along a temperature gradient, providing control over microparticles~\cite{liu2021opto}; active optothermophoretic manipulation involves particles that can establish their own temperature gradients~\cite{liu2021opto}.

With varied possibilities for driving mechanisms, fluid flow is instrumental in many techniques for micromanipulation. 
One of its advantages is that it need not rely on the optical or magnetic properties of the particles being manipulated; instead, recent studies have employed a vast range of flow field topologies in order to trap or manoeuvre particles hydrodynamically. 
For instance, time-averaged flow created by fluid oscillation, known as steady streaming, has enabled contact-free trapping of single cells in microeddies~\cite{baudoin2020acoustic,lutz2006hydrodynamic}.
Both driven by laser-induced temperature gradients, thermal convection and thermophoresis have been combined to manipulate microparticles~\cite{qian2020microparticle}, while Marangoni flows can be controlled optothermally~\cite{winterer2018optofluidic} or via photoresponsive surfactants~\cite{varanakkottu2013particle}, to trap and guide the movement of particles at a gas-liquid interface, or to transport living cells~\cite{hu2013opto}.
Electroosmotic flow, which is actuated by electrodes that generate electric fields and hence flow of electrolyte solutions, has been used together with feedback control to independently steer and trap multiple particles at once~\cite{armani2006using}.
In a microfluidic device with multiple channels, pressure-driven flow can be controlled via the channel flow rates; using this flow control, recent experiments have demonstrated precise manipulation of particles along user-defined paths, exploiting different flow modes~\cite{tanyeri2010hydrodynamic,shenoy2016stokes,schneider2011algorithm,tu20233d,gonzalez2024symmetry}.
However, increasing the number of particles increases the number of degrees of freedom, necessitating a larger number of channels of the microfluidic device to control the particles independently~\cite{schneider2011algorithm}.

Recent microfluidic experiments have investigated the use of thermoviscous flows for high-precision positioning of particles~\cite{erben2021feedback}, trapping~\cite{stoev2021highly} and assembly~\cite{erben2024opto}. 
In these studies, thermoviscous flow is driven by spatio-temporally varying heating of fluid, induced by a laser scanning along specified paths at kilohertz frequency. 
Although this technique, too, involves laser-induced heating, the physical mechanism for flow distinguishes it from the methods previously described. Here, the repeated scanning of the laser along a scan path results in fluid flow through the interplay between thermal expansion and thermal viscosity changes. This enables localised net transport of particles in the bulk fluid, without requiring direct laser exposure or physical channels~\cite{weinert2008optically,erben2024opto}. 

Highly versatile, these thermoviscous flows have also been applied inside living cells, and are known as focused-light-induced cytoplasmic streaming (FLUCS) in this context~\cite{mittasch2018non,seelbinder2024probe,minopoli2023iso}. This name contrasts FLUCS with naturally occurring cytoplasmic streaming: actively driven flows of the water-based, complex fluid (cytoplasm) inside cells, found in a wide variety of living organisms~\cite{lu2023go,goldstein2015physical}. 
For example, with thermoviscous flow perturbations, experiments have investigated how intracellular flows drive cell polarisation~\cite{mittasch2018non}, and measured the rheology of cytoplasm~\cite{mittasch2018non} and the nucleus~\cite{seelbinder2024probe}. 

\begin{figure}[t]
	\centering
	{\includegraphics[width=0.8\textwidth]{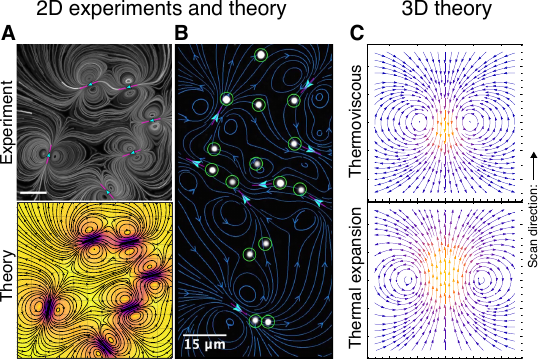}
		\caption{Two-dimensional (A, B) and three-dimensional (C) thermoviscous and thermal expansion-driven net flows.
				(A) Trajectories of tracers in viscous fluid confined between parallel plates, induced by repeated scanning of six scan paths, in an experiment~\cite{erben2024opto} (top) and according to analytical modelling of net thermoviscous flows~\cite{liao2023theoretical,erben2024opto} (bottom). Scale bar: $15~\si{\micro\metre}$. 
				(B) Control of 15 microparticles (white) with 8 scan paths (magenta) to form a humanoid figure in an experiment~\cite{erben2024opto}. 
				The net thermoviscous flow predicted by theory~\cite{liao2023theoretical,erben2024opto} is shown in blue, with target positions in green. Scale bar: $15~\si{\micro\metre}$.
				Panels~A~and~B adapted from Ref.~\cite{erben2024opto} and licensed under \href{https://creativecommons.org/licenses/by/4.0/}{CC BY 4.0}.
				(C) Theoretical trajectories of tracers in three-dimensional, unconfined fluid induced by scanning of a spherical heat spot (scan direction indicated by arrow), as derived in this work, due to thermoviscous effects (top) and due to purely thermal expansion-driven flows (bottom).
		}
		\label{fig:motivating_axisymmetric_v1_05022025}}
\end{figure}

Previous theoretical work~\cite{liao2023theoretical,weinert2008microscale,weinert2008optically} has focused on the setup relevant to many of these microfluidic experiments~\cite{liao2023theoretical,erben2021feedback,minopoli2023iso,erben2024opto,stoev2021highly,mittasch2018non,weinert2008microscale,weinert2008optically}, namely, with viscous fluid confined between parallel rigid plates.
In Ref.~\cite{liao2023theoretical}, we considered a temperature profile with circular symmetry induced by the laser, which translates along a finite scan path, always in the same direction. 
We modelled the density and shear viscosity of the fluid as depending linearly on temperature change, a valid approximation for experimentally relevant temperature increases. 
The confinement provided by the parallel plates reduced the problem to two spatial dimensions, with a parabolic flow profile along the third dimension in the lubrication limit.
We solved for the instantaneous flow field, and hence the average velocity of tracer particles throughout space due to repeated scanning.
This revealed strong net transport near the scan path, together with inverse-square-law spatial decay, in agreement with experimental data~\cite{erben2021feedback}.
Perhaps counter-intuitively, this net transport of tracers depends not on the magnitude, but instead on the rate of change of shear viscosity with temperature; hence, net thermoviscous flows are effective in fluids with vastly different viscosities~\cite{mittasch2018non,liao2023theoretical,weinert2008microscale}. 
Extending the method from one scan path to many, feedback algorithms have enabled the simultaneous manipulation of multiple particles, via net flow fields with complex topology~\cite{minopoli2023iso,erben2024opto,erben2021feedback}, as shown in Fig.~\ref{fig:motivating_axisymmetric_v1_05022025}A~and~B.

In general, micromanipulation can be conducted in two ~\cite{de2012recent,liu2021opto,qian2020microparticle,tanyeri2010hydrodynamic,shenoy2016stokes,schneider2011algorithm} or three~\cite{moffitt2008recent,de2012recent,gosse2002magnetic,baudoin2020acoustic,liu2021opto,tu20233d,gonzalez2024symmetry} dimensions. 
So far, thermoviscous flows have been used for two-dimensional micromanipulation experimentally~\cite{weinert2008optically,erben2021feedback,stoev2021highly,erben2024opto}, while existing theoretical studies consider essentially two-dimensional flow in the parallel-plate geometry~\cite{liao2023theoretical,weinert2008microscale}. 
However, the possibility of three-dimensional thermoviscous and thermal expansion-driven net flows and micromanipulation, for example, with highly focused heating~\cite{weinert2008optically,liao2023theoretical}, has yet to be explored. 
Unlocking more degrees of freedom could result in increased versatility of the technique for future experiments.

Geometry has a significant influence on the flows and transport induced by the focused light. 
For example, it can strongly modify the far-field behaviour, which, importantly, determines the impact of directing a particle towards its target on other particles, in microfluidic experiments~\cite{erben2024opto}. 
Thus, to make progress towards three-dimensional micromanipulation, fundamental, quantitative understanding of three-dimensional flow is required.
In this article, we study thermoviscous and thermal expansion-driven flows of three-dimensional, unconfined fluid.

As a consequence of the unconfined geometry, we must also include a qualitatively new physical ingredient, known as the bulk viscosity (or volume viscosity~\cite{happel}) of the fluid, in our theory.
This material property is relevant only for compressible flows. 
For the parallel-plate geometry, it  could be safely neglected~\cite{liao2023theoretical,weinert2008microscale}; however, without the parallel plates, bulk viscosity could play a role. 
From a continuum perspective, just like shear viscosity, bulk viscosity is a phenomenological coefficient that characterises a Newtonian fluid, and is independent of its state of motion~\cite{happel}.
While shear viscosity relates stress to linear deformation rate, bulk viscosity relates stress to volumetric deformation rate~\cite{happel}, i.e.~dilatational-compressional motion of the fluid described by divergence of flow velocity~\cite{holmes2011temperature}.
At the molecular level, bulk viscosity reflects rotational and vibrational degrees of freedom in molecular motion, while shear viscosity is associated with translational motion~\cite{temkin1981elements,dukhin2009bulk}.

The bulk viscosity of a liquid may be determined experimentally via various techniques~\cite{slie1966ultrasonic,dukhin2009bulk,holmes2011temperature,xu2003measurement}.  
Motivated by recent microfluidic thermoviscous flow experiments~\cite{mittasch2018non,weinert2008optically,weinert2008microscale}, we may focus here on both glycerol-water mixtures and pure water. 
The authors of Ref.~\cite{slie1966ultrasonic} demonstrated that the bulk viscosity of aqueous glycerol is comparable in magnitude with the shear viscosity, and that it depends on temperature.
Furthermore, different groups have shown that the bulk viscosity of water decreases with temperature, at a rate comparable to that of shear viscosity, over the temperature range relevant to thermoviscous flow experiments~\cite{slie1966ultrasonic,holmes2011temperature,xu2003measurement}.
Therefore, we cannot neglect the magnitude and rate of change with temperature of bulk viscosity in favour of those of shear viscosity, and here we retain both bulk viscosity and its temperature dependence in our theory.

In this paper, we solve systematically for the flow and net transport of particles that result from the scanning of a model spherical heat spot in unconfined viscous fluid. 
This article is organised as follows. 
We begin in Sec.~\ref{sec:temperature} with the heat transport problem, solving numerically for the temperature field induced by a prescribed translating heat source in the experimentally relevant limit. Based on our results, we then choose a temperature profile that will serve as an input to our fluid flow model, representing the effect of the spatio-temporally varying heating.
Next, in Sec.~\ref{sec:flow}, we introduce our model for the compressible fluid flow, including the effects of thermal expansion, thermal shear viscosity changes, and bulk viscosity with arbitrary temperature-dependence. 
We derive an analytical expression for the flow induced during the translation of the heat spot up to quadratic order.
Although our model includes the effect of bulk viscosity, we demonstrate that the flow is independent of it, instead depending on thermal expansion and thermal shear viscosity changes, just as for the parallel-plate geometry.
We then compute in Sec.~\ref{sec:transport} the net transport of passive tracers induced by the flow, due to a full scan of the heat spot. 
In contrast with the instantaneous fluid flow, the leading-order net transport occurs at quadratic order in the thermal expansion coefficient and thermal shear viscosity coefficient.
The direction of the net transport generated depends on the relative importance of these two effects, which can provide competing contributions: one from the interplay of thermal expansion and thermal shear viscosity changes (thermoviscous net transport, shown in the top panel of Fig.~\ref{fig:motivating_axisymmetric_v1_05022025}C), and the other purely driven by thermal expansion (Fig.~\ref{fig:motivating_axisymmetric_v1_05022025}C, bottom).
Furthermore, the average velocity of tracers decays as a hydrodynamic source dipole in the far field, more strongly than the instantaneous flow during one scan of the heat spot.
Finally, in Sec.~\ref{sec:discussion}, we compare the results of our axisymmetric model with our previous work for the parallel-plate setup, and propose future experiments to validate our theory, along with potential applications.

\section{Temperature field induced by scanning heat source}\label{sec:temperature}

In this section, we solve numerically for the temperature field induced by a prescribed heat source that moves along a scan path, in three-dimensional, unbounded space, in the limit relevant experimentally to FLUCS~\cite{mittasch2018non,minopoli2023iso}. 
Motivated by our results, we then introduce the temperature profile that we will prescribe in our model of the flow induced by the scanning heat source (Sec.~\ref{sec:flow}). 

\subsection{Setup}

\begin{figure}[t]
		\centering
	{\includegraphics[width=0.25\textwidth]{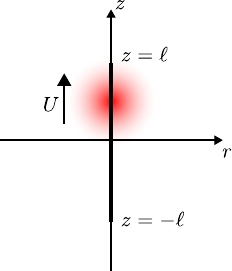}
		\caption{Setup for our model of heat transport induced by a scanning laser. A spherical heat source of characteristic radius $b$ translates at speed $U$ in the $z$ direction, along a scan path from $z=-\ell$ to $z=\ell$ along the $z$ axis (cylindrical radial coordinate $r$), in unbounded, viscous fluid, causing a localised temperature perturbation.}
		\label{fig:fig_diagram_setup_v1_11082024}}
\end{figure}

We illustrate the setup in Fig.~\ref{fig:fig_diagram_setup_v1_11082024}, as a model for the standard FLUCS setup~\cite{mittasch2018non}.  
A prescribed heat source of characteristic radius $b$ translates at constant speed $U$, along a scan path on the $z$ axis from $z=-\ell$ to $z=\ell$, in three-dimensional, unconfined fluid. 
The setup is axisymmetric about the $z$ axis; we denote the cylindrical radial coordinate by $r$.
We consider one scan of the heat source: the scan starts at time $t=-t_0$ and finishes at time $t=t_0$, so that the centre of the heat source has position $(r=0, z=Ut)$ at time $t$, and the scan period is given by $2t_0\equiv 2\ell/U$.
This heat source induces a localised temperature increase in the fluid, which in turn drives fluid flow (modelled in Sec.~\ref{sec:flow}).

\subsection{Governing equation for heat transport}

We consider the heat transport problem that determines the temperature field $T$ due to the prescribed heat source.
We first make the simplifying assumption to neglect advection of heat by fluid flow: it may be verified \textit{a posteriori} (Sec.~\ref{sec:beta_n2_gamma_n3}) that the characteristic scale for the term $\partial T/\partial t$ is much larger than for the advective term $\mathbf{u}\cdot\nabla T$~\cite{liao2023theoretical}, where $\mathbf{u}$ is the fluid velocity field, under the same assumptions we employ to solve for flow; 
the key is that asymptotically, the flow velocity we will obtain in Sec.~\ref{sec:flow} is much smaller than the translation speed of the heat source. 
The temperature field therefore evolves primarily because of thermal diffusion and the prescribed heat source, independently of the flow driven by the temperature perturbation.
Further, as a first approximation in this section on heat transport, we treat the material properties of the fluid as constant; however, when we consider the fluid flow driven by the temperature field in Sec.~\ref{sec:flow}, we will allow the material properties to vary with temperature.
Thus, the temperature field is governed by the forced heat equation as
\begin{align}
	\rho_0 c_p \frac{\partial T}{\partial t}  = k \nabla^2 T + \Phi,
\end{align}
where $\rho_0$, $c_p$ and $k$ are the (constant) density, specific heat capacity and thermal conductivity of the fluid, respectively, and $\Phi$ is the source term, which provides the forcing. 
For the source term, we prescribe a translating, spherically symmetrical, Gaussian profile with time-dependent (non-negative) amplitude as 
\begin{align}
	\Phi = \frac{\Phi_0}{\sqrt{2\pi}} B(t) \exp\{- [r^2 + (z-Ut)^2] /2b^2\},
\end{align}
where $\Phi_0$ is the characteristic scale for the heat source term and the dimensionless amplitude function $B(t)$ is given by
\begin{align}
	B(t) = \begin{cases}
		\cos\left (\frac{\pi t}{2t_0}\right )^2  &\text{for $-t_0 \leq t \leq t_0$}, \\
		0 & \text{otherwise}. \label{eq:amplitude_B}
	\end{cases}
\end{align}
Highly focused heating of the fluid has been suggested as a potential method of achieving this experimentally~\cite{weinert2008optically,mittasch2018non}. 
Here we consider only one scan, after which the heat source switches off completely, to investigate the temperature field (we will consider the effect of repeated scanning on net fluid flow in later sections). 
For convenience, we may write the temperature field $T(r,z,t)$ of the fluid as
\begin{align}
	T(r,z,t) = T_0 + \Delta T(r,z,t),
\end{align}
where $T_0$ is a constant reference temperature and $\Delta T(r,z,t)$ is the temperature change of the fluid due to the localised heat source, which we assume to decay to zero at infinity.

\subsection{Nondimensionalisation}

For the temperature problem, we nondimensionalise length with $b$, time with $b/U$ and temperature change with $\Phi_0 b^2/k$; we note that for the later sections on fluid flow (Sec.~\ref{sec:flow} and Sec.~\ref{sec:transport}), we will use a different nondimensionalisation. 
In what follows, we use variable names to mean their dimensionless equivalents, to simplify notation.
The dimensionless forced heat equation then becomes
\begin{align}
	\text{Pe}_\text{scan} \frac{\partial \Delta T}{\partial t} =  \nabla^2 \Delta T + \Phi,\label{eq:heat_equation}
\end{align}
where the dimensionless heat source term $\Phi$ is given by
\begin{align}
	\Phi = \frac{1}{\sqrt{2\pi}} B(t) \exp\{- [r^2 + (z-t)^2] /2\},\label{eq:Phi_dimensionless}
\end{align}
and we define the parameter $\text{Pe}_\text{scan}$ to be the scanning P\'eclet number, given by
\begin{align}
	\text{Pe}_\text{scan} \equiv \frac{U b}{(k / \rho_0 c_p)},
\end{align}
where the denominator is the thermal diffusivity. 
$\text{Pe}_\text{scan} $ is a dimensionless ratio that quantifies the relative importance of the scanning speed and thermal diffusion.
It may be thought of as an unsteady P\'eclet number; importantly, the characteristic speed involved is the scanning speed, and not the fluid flow speed that would feature in the classical P\'eclet number.

\subsection{Numerical simulation details}\label{sec:temp_parameters}

To obtain an estimate for $\text{Pe}_\text{scan}$, we substitute parameter values for water at $20~\si{\celsius}$ at atmospheric pressure~\cite{dinccer2016drying}, given by $k=0.5861~\si{\watt}~\si{\metre}^{-1}~\si{\kelvin}^{-1}$, $\rho_0=998.2~\si{\kilogram}~\si{\metre}^{-3}$ and $c_p = 4183~\si{\joule}~\si{\kilogram}^{-1}~\si{\kelvin}^{-1}$.
Although the geometry is new in the present work, we use sample parameter values (dimensional) for the scan properties based on previous  FLUCS and microfluidic experiments in confined geometries~\cite{mittasch2018non,erben2021feedback}, setting the heat source radius as $b=4~\si{\micro\metre}$ and the scan path length as $2\ell=11~\si{\micro\metre}$, so that the dimensionless half-scan period is given by $t_0=1.375$.
With these parameter values and selected dimensional scan periods, we solve the dimensionless heat equation [Eq.~\eqref{eq:heat_equation}] numerically, using finite element analysis in MATLAB R2024a. 
We present results for two different scanning P\'eclet numbers: first, $\text{Pe}_\text{scan}=0.63$, which corresponds to scan frequency $2~\si{\kilo\hertz}$ (i.e.~dimensional scan period $2t_0 = 0.5~\si{\milli\second}$ and speed $U=0.022~\si{\metre}~\si{\second}^{-1}$), typical for FLUCS experiments~\cite{mittasch2018non,minopoli2023iso,erben2021feedback}; then $\text{Pe}_\text{scan}=2$, representing a faster scan within the experimental range.

\subsection{Results from numerical simulation of forced heat equation}

\subsubsection{Slower scan ($\text{Pe}_\text{scan}=0.63$)}

\begin{figure}[t]
		\centering
	{\includegraphics[width=0.9\textwidth]{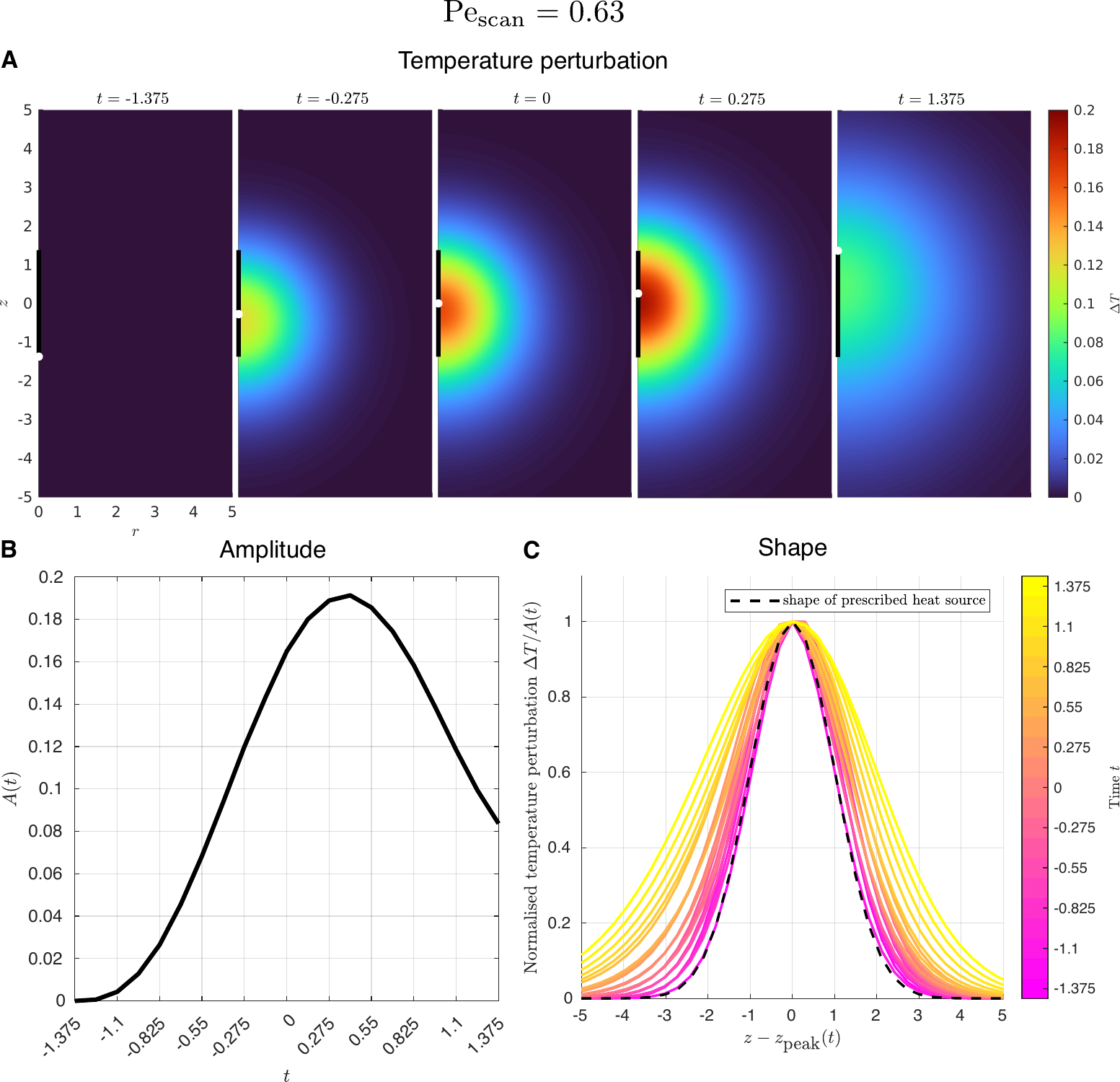}
		\caption{Temperature profile for numerical simulations of forced heat equation for scanning P\'eclet number $\text{Pe}_\text{scan}=0.63$, during one scan of the heat source ($-1.375\leq t \leq 1.375$). 	
			(A) Heat map showing spatial variation of temperature perturbation $\Delta T$ with cylindrical  coordinates $r$ and $z$, at selected times. The centre of the heat source is indicated in white, with the scan path in black.
			(B) Amplitude of temperature perturbation $A(t)$ as a function of time.
			(C) Shape of temperature perturbation $\Delta T/A(t)$, i.e. temperature perturbation normalised by its peak, along the $z$ axis, at selected times. The location of the peak temperature on the $z$ axis is given by $z=z_\text{peak}(t)$. 
			Colours change from pink to yellow as time $t$ increases. 
			The dashed curve indicates the shape of the prescribed heat source~[Eq.~\eqref{eq:analytical_shape}], for comparison.
		}
		\label{fig:plots_numerics_unsteady_diffusion_translating_on_off_source_v6_16012025}}
\end{figure}

First, for the slower scan of the heat source with scanning P\'eclet number $\text{Pe}_\text{scan}=0.63$ (parameter values in Sec.~\ref{sec:temp_parameters}), we illustrate in Fig.~\ref{fig:plots_numerics_unsteady_diffusion_translating_on_off_source_v6_16012025} the temperature field induced during one scan period ($-1.375\leq t \leq 1.375$), as obtained by numerical simulation of the forced heat equation.
In Fig.~\ref{fig:plots_numerics_unsteady_diffusion_translating_on_off_source_v6_16012025}A, we show snapshots of the temperature field as the heat source translates along the scan path.
We observe that the region of higher temperature is highly localised and approximately spherically symmetrical. 
Like the prescribed heat source, this temperature perturbation translates in the positive $z$ direction, first increasing in amplitude, peaking during the scan period and then decreasing.
Any heat that remains at the end of the scan of the heat source diffuses away afterwards, so the temperature decays to ambient.

To further characterise the temperature field, we plot in Fig.~\ref{fig:plots_numerics_unsteady_diffusion_translating_on_off_source_v6_16012025}B and Fig.~\ref{fig:plots_numerics_unsteady_diffusion_translating_on_off_source_v6_16012025}C the amplitude and the shape of the temperature perturbation, respectively. 
Here, we define the amplitude $A(t)$ (dimensionless) as the instantaneous maximum temperature perturbation at time $t$, which occurs at position $z=z_\text{peak}(t)$ on the $z$ axis, while the shape is the temperature perturbation field $\Delta T$ divided by this amplitude, given by $\Delta T/A(t)$.
Over the course of a scan period, the amplitude (Fig.~\ref{fig:plots_numerics_unsteady_diffusion_translating_on_off_source_v6_16012025}B) increases, peaks just after halfway through the scan period, and then decreases.
Note also that the amplitude is not precisely zero at the end of the scan period, as it takes time for the heat remaining after the heat source has been switched off to diffuse away.
The amplitude of the temperature perturbation therefore inherits broad qualitative features from that of the heat source, with differences in finer details.

In Fig.~\ref{fig:plots_numerics_unsteady_diffusion_translating_on_off_source_v6_16012025}C, we plot the shape of the temperature perturbation along the $z$ axis, as a function of displacement $z-z_\text{peak}(t)$ from the location of the instantaneous maximum temperature, at selected times.
For comparison, the black dashed curve is given by the formula
\begin{align}
	\frac{\Delta T}{A(t)} = \exp\{-[z-z_\text{peak}(t)]^2/2\}.\label{eq:analytical_shape}
\end{align}
This is a Gaussian of the same radius as the prescribed heat source (centred at $z_\text{peak}(t)$).
The key observation here is that strikingly, the shape is well-approximated by a Gaussian throughout the scan.
As time progresses, the shape exhibits a small increase in radius due to diffusive spreading.
The characteristic distance by which the temperature perturbation spreads out can be determined via an intuitive scaling argument. 
In the (dimensional) scan period $2t_0$, heat diffuses by a characteristic distance of $\sqrt {(k/\rho_0 c_p) (2t_0)} \approx 8~\si{\micro\metre}$ for our parameter values, which corresponds to a dimensionless distance of approximately $2$, in good agreement with the degree of spreading seen in  Fig.~\ref{fig:plots_numerics_unsteady_diffusion_translating_on_off_source_v6_16012025}C.

\subsubsection{Faster scan ($\text{Pe}_\text{scan}=2$)}

\begin{figure}[t]
		\centering
	{\includegraphics[width=0.5\textwidth]{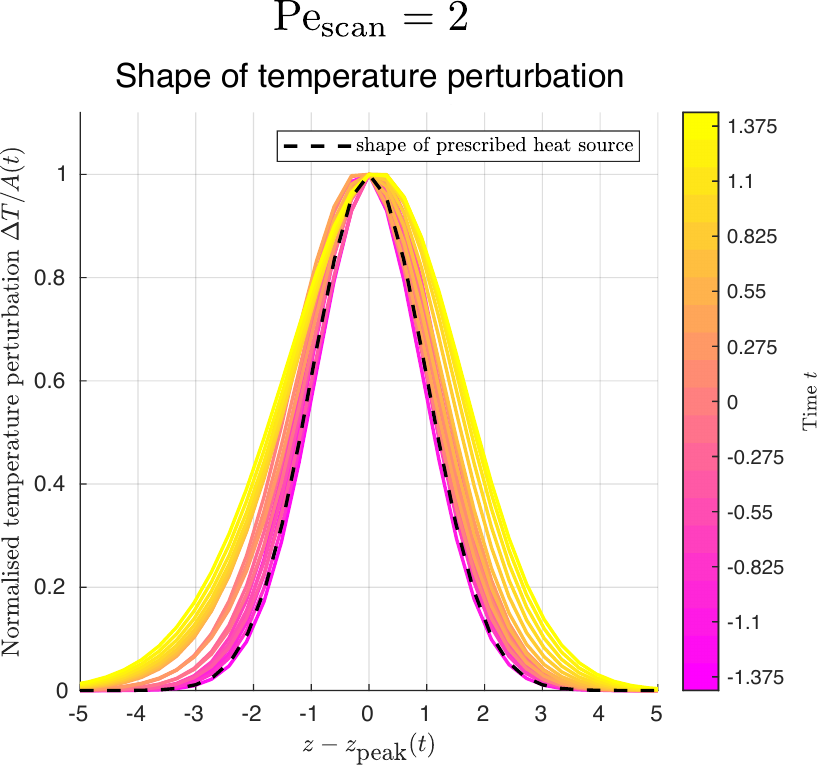}
		\caption{Shape of temperature perturbation $\Delta T/A(t)$ along the $z$ axis, at selected times, as obtained by numerical simulation of the forced heat equation for scanning P\'eclet number~$\text{Pe}_\text{scan}=2$, during one scan of the heat source ($-1.375\leq t \leq 1.375$), for comparison with result for $\text{Pe}_\text{scan}=0.63$ in Fig.~\ref{fig:plots_numerics_unsteady_diffusion_translating_on_off_source_v6_16012025}C. 
			The dashed curve shows the shape of the prescribed heat source~[Eq.~\eqref{eq:analytical_shape}].
		}
		\label{fig:plots_numerics_higher_Pe_unsteady_diffusion_translating_on_off_source_v6_16012025}}
\end{figure}

To explore the diffusive spreading of the Gaussian shape further, we now consider a higher scanning P\'eclet number (faster scanning). 
For $\text{Pe}_\text{scan}=2$, we illustrate the shape of the temperature perturbation $\Delta T/A(t)$ in Fig.~\ref{fig:plots_numerics_higher_Pe_unsteady_diffusion_translating_on_off_source_v6_16012025}.
We see reduced spreading in this case. 
Indeed, in the limit of high $\text{Pe}_\text{scan}$, completely neglecting the Laplacian (diffusion) term on the right-hand side of the heat equation [Eq.~\eqref{eq:heat_equation}] allows us to integrate once with respect to time and hence write down the temperature field in this regime as
\begin{align}
	\Delta T = \frac{1}{\sqrt{2\pi}\text{Pe}_\text{scan}} \exp(-r^2/2) \int_{-t_0}^{t} B(\tilde{t})\exp[-(z-\tilde{t})^2/2] \, d\tilde{t}.\label{eq:temperature_no_diffusion}
\end{align}
It may be verified that this essentially inherits the Gaussian shape and radius of the prescribed heat source (by inspection for the $r$ dependence and by considering the far field for the $z$ dependence).

Therefore, the initial Gaussian shape and radius originate from the heat source directly setting the evolution of the temperature field; during the scan, the shape remains Gaussian and the radius increases a little due to diffusion.
Crucially, from a mathematical perspective, the unsteady term $\text{Pe}_\text{scan} \frac{\partial \Delta T}{\partial t}$ in Eq.~\eqref{eq:heat_equation} must be retained. 
This is confirmed in the case of the two-dimensional (parallel-plate) geometry by experimental measurements showing that the temperature perturbation is highly localised~\cite{weinert2008optically,mittasch2018non}, an observation only consistent with the forced diffusion equation when the unsteady term is not discarded.

\subsection{Model for temperature profile}\label{sec:temperature_model}

We are now ready to propose a simplified temperature profile, as an input for our fluid flow model in the next section. 
To model the effect of the heating on the flow during one scan, we will prescribe a Gaussian temperature perturbation with time-varying amplitude, given dimensionally by
\begin{align}
	\Delta T(r,z,t) = \Delta T_0 A(t)\exp\{-[r^2+(z-Ut)^2]/2a^2\},\label{eq:Gaussian_temp}
\end{align}
where $\Delta T_0$ is the characteristic temperature change (a positive constant), $A(t)$ is a dimensionless amplitude function (taking non-negative values), $U$ is the speed of translation and $a$ is the characteristic radius.
Here, the amplitude function can be thought of as arbitrary or prescribed.
This temperature profile captures the essential features of the solution to the heat transport problem we solved numerically, needed to explain thermal expansion-driven and thermoviscous fluid flows: the translation, shape and time-dependent amplitude.
In particular, we have approximated the radius of the Gaussian as constant, neglecting the limited amount of diffusive spreading observed in the numerical simulations; in two dimensions, the thermoviscous net flows calculated theoretically under the same assumption have been shown to agree well with experimental results~\cite{liao2023theoretical,erben2024opto}.

In our modelling in Eq.~\eqref{eq:Gaussian_temp}, we need not select the same value for the characteristic radius of the temperature perturbation $a$ as for that of the heat source $b$, although we may choose to as a modelling assumption.
Similar applies for the amplitude function.
We note also that we will later assume that the amplitude of the temperature perturbation is zero at the start and end of a scan, following Ref.~\cite{liao2023theoretical}. 
This is a simplifying approximation, as we saw that the temperature decays to ambient over time at the end of a scan by diffusion, according to the heat equation.
This approximation will allow us to treat every scan in repeated scanning as the same, as we assume that the temperature field resets to ambient at the end of every scan.

Finally, we recall that in the parallel-plate geometry, the temperature profile was also modelled as Gaussian~\cite{liao2023theoretical,weinert2008optically} based on experiments~\cite{weinert2008optically,mittasch2018non}, with cylindrical instead of spherical symmetry. 
Our analogous temperature profile for three-dimensional geometry will enable direct comparison of our results with the two-dimensional case.

\section{Instantaneous flow during one scan}\label{sec:flow}

In this section, we introduce a three-dimensional model for the thermoviscous and thermal expansion-driven flows induced during one scan of a translating heat spot, in unbounded fluid.
Here and in what follows, the term ``heat spot" refers to the model temperature perturbation detailed in Sec.~\ref{sec:temperature_model}.
With the instantaneous flow during one scan, we will compute in Sec.~\ref{sec:transport} the net transport of particles due to a full scan, which, by analogy with the two-dimensional case~\cite{liao2023theoretical}, we anticipate will allow trapping and manipulation of micron-sized particles in future experiments.

\subsection{Setup}

The setup is essentially as illustrated in Fig.~\ref{fig:fig_diagram_setup_v1_11082024}.
We now consider a localised, spherically symmetric temperature perturbation of characteristic radius $a$ that translates along a scan path, given by the line segment from $z=-\ell$ to $z=\ell$ along the $z$ axis, at speed $U$ in unbounded, viscous fluid.
During the scan, the centre of the heat spot is therefore at $(r=0, z=Ut)$ at time $t$, for~$-t_0 \leq t \leq t_0$, where the scan period is $2t_0 \equiv 2\ell/U$; the heat spot travels upwards. 
Again, the setup is axisymmetric about the $z$ axis, with cylindrical radial coordinate denoted by $r$.
We aim in this section to solve for the instantaneous fluid flow during one scan, driven by thermal expansion and thermal viscosity changes; we will build on this in Sec.~\ref{sec:transport} to understand the net transport of tracers due to repeated scanning of the heat spot, always upwards. 

\subsection{Governing equations}

The translating localised temperature increase in the fluid modifies its material properties locally, driving fluid flow governed by the mass conservation and the momentum equations.
Most of the explicit formulae for flow derived in this article correspond to a prescribed Gaussian temperature perturbation $\Delta T(r,z,t)$ with time-varying dimensionless amplitude $A(t)$, introduced in Sec.~\ref{sec:temperature} and given by Eq.~\eqref{eq:Gaussian_temp}.
This model will allow us to understand the physics of thermoviscous and thermal expansion-driven flows.
For small temperature changes, we use a standard linear relationship between the density of the fluid and the temperature, given by
\begin{align}
	\rho &= \rho_0 (1 - \alpha \Delta T),\label{eq:density_temp_diml_axisym}
\end{align}
where $\rho_0$ is the density of the fluid at the reference temperature $T_0$, and $\alpha$ is the thermal expansion coefficient (typically positive).

The mass conservation equation is given by
\begin{align}
	\frac{\partial \rho}{\partial t} + \nabla\cdot(\rho\mathbf{u})=0,\label{eq:mass_consn_3D_axisym}
\end{align}
where $\mathbf{u}$ is the velocity field; thus, the spatio-temporally varying density field gives rise to compressible fluid flow.
The Cauchy momentum equation is given by
\begin{align}
	\rho \frac{D \mathbf{u}}{Dt} = \nabla \cdot \boldsymbol{\Pi} + \rho \mathbf{g},\label{eq:Cauchy}
\end{align}
where $\mathbf{g}$ is the gravitational acceleration and the stress tensor $\boldsymbol{\Pi}$ is given by 
\begin{align}
	\boldsymbol{\Pi} = - p \mathbf{1} + \kappa (\nabla \cdot \mathbf{u}) \mathbf{1} 
	+ 2 \eta \left \{ \frac{1}{2} [ \nabla \mathbf{u} + (\nabla\mathbf{u})^\text{T} ] - \frac{1}{3} \mathbf{1} (\nabla \cdot \mathbf{u} ) \right \},\label{eq:Pi}
\end{align}
where $p$  is the pressure field, $\mathbf{1}$ is the identity tensor, and  $\eta$ and $\kappa$ are, respectively, the shear and bulk viscosities of the fluid.
In this work, we consider a regime in which inertia and gravity are both negligible, relevant at small length scales; the dimensionless numbers associated with these conditions may be found by scaling arguments similar to those presented in Ref.~\cite{liao2023theoretical}.
For example, adapting to our geometry, the gravity-driven flow (natural convection) scales as $\rho_0 g a^2 \alpha \Delta T_0 / \eta_0$, whereas the thermal expansion-driven flow scales as $\alpha \Delta T_0 U$.
Thus, natural convection may be neglected if the dimensionless ratio $\rho_0 g a^2/\eta_0 U$ is small; we calculate its value as less than 0.01, for water at $20~\si{\celsius}$ at atmospheric pressure~\cite{dinccer2016drying} and estimated parameter values of heat-spot radius $a=4~\si{\micro\metre}$ and speed $U=0.022~\si{\metre}~\si{\second}^{-1}$~\cite{mittasch2018non,erben2021feedback}.

The shear viscosity $\eta$ varies with temperature of the fluid, which we model (as in previous studies~\cite{liao2023theoretical,weinert2008microscale,weinert2008optically}) as
\begin{align}
	\eta &= \eta_0 (1 - \beta \Delta T),\label{eq:viscosity_temp_diml_axisym}
\end{align}
where $\eta_0$ is the shear viscosity of the fluid at the reference temperature $T_0$, and $\beta$ is the thermal shear viscosity coefficient (typically positive). 

We recall that in the lubrication limit for the parallel-plate setup, the effect of bulk viscosity was shown to be unimportant~\cite{weinert2008microscale,liao2023theoretical}. 
However, for aqueous glycerol, shear viscosity and bulk viscosity are similar in magnitude~\cite{slie1966ultrasonic,dukhin2009bulk,holmes2011temperature}; furthermore, bulk viscosity varies with temperature~\cite{slie1966ultrasonic,holmes2011temperature,xu2003measurement}.
Thus, in our theory, we must retain the bulk viscosity. 
It would be possible to treat the temperature dependence of bulk viscosity in the same way as for density or shear viscosity for small temperature changes.
That is, we could model the bulk viscosity as depending linearly on temperature, introducing a parameter to characterise the rate of variation.
However, here, this turns out to be unnecessary and, as we show below, we can instead treat the bulk viscosity in full generality, without assuming a specific functional form for its dependence on temperature.
We finally note that for a temperature perturbation that decays at infinity, the bulk viscosity, correspondingly, tends to a constant at infinity, just as the fluid density and shear viscosity do.

\subsection{Nondimensionalisation}\label{sec:nondiml}

We nondimensionalise length with $a$, velocity with~$U$, time with~$a/U$, pressure with the viscous scale~$\eta_0 U/a$, both shear viscosity and bulk viscosity with~$\eta_0$, density with~$\rho_0$ and temperature with~$\Delta T_0$; note that this nondimensionalisation differs to that for the thermal problem in Sec.~\ref{sec:temperature}. 
In what follows, we use variable names to mean their dimensionless versions for notational convenience.  
We summarise the dimensionless governing equations.
The momentum equation [Eqs.~\eqref{eq:Cauchy}--\eqref{eq:Pi}] becomes
\begin{align}
	- \nabla p 
	+ \nabla (\kappa \nabla \cdot \mathbf{u})
	+ \eta \nabla^2 \mathbf{u} + \frac{1}{3} \eta \nabla (\nabla \cdot \mathbf{u})
	+ (\nabla \eta) \cdot [\nabla \mathbf{u} + (\nabla \mathbf{u})^\text{T}] - \frac{2}{3} (\nabla \eta) (\nabla \cdot \mathbf{u})
	&= \mathbf{0}.\label{eq:momentum}
\end{align}
Here, we draw attention to the bulk viscosity term $\nabla (\kappa \nabla \cdot \mathbf{u})$, which appears in the momentum equation as an exact gradient.
Thus, a solution $(\mathbf{u}',p')$ to the governing equations with bulk viscosity set to zero induces a solution $(\mathbf{u},p)$ to the full equations with arbitrary bulk viscosity, with the two pressure fields related via $p'=p-\kappa \nabla \cdot \mathbf{u}$.

After nondimensionalisation, the mass conservation equation [Eq.~\eqref{eq:mass_consn_3D_axisym}] is still given by
\begin{align}
 	\frac{\partial \rho}{\partial t} + \nabla\cdot(\rho\mathbf{u})=0.\label{eq:mass_consn_3D_axisym_dimless}
\end{align}
The density [Eq.~\eqref{eq:density_temp_diml_axisym}] and shear viscosity [Eq.~\eqref{eq:viscosity_temp_diml_axisym}] of the fluid are now related to the temperature change by 
\begin{align}
 	\rho &= 1 - \alpha \Delta T,\label{eq:density_temp_dimless_axisym}
\end{align}
and
\begin{align}
 	\eta &= 1 - \beta \Delta T,\label{eq:viscosity_temp_dimless_axisym}
\end{align}
respectively, while the Gaussian model temperature perturbation [from Eq.~\eqref{eq:Gaussian_temp}] becomes
\begin{align}
 	\Delta T(r,z,t) = A(t)\exp\{-[r^2+(z-t)^2]/2\}.\label{eq:Gaussian_temp_dimless}
\end{align}

\subsection{Perturbation expansion}

Due to our assumption of small temperature changes, the dimensionless coefficients $\alpha$ and $\beta$ are now small parameters, representing the proportion by which the density and shear viscosity decrease in response to a temperature increase of $\Delta T_0$, respectively.
We therefore solve this problem perturbatively, with the boundary conditions that the velocity and pressure are non-singular at the origin and decay at infinity (for temperature profiles that decay at infinity). 
We pose perturbation expansions for the velocity field and pressure field in the two small parameters, given by
\begin{align}
	\mathbf{u} &= \mathbf{u}_{0,0} + \alpha \mathbf{u}_{1,0} + \beta \mathbf{u}_{0,1} 
	+ \alpha^2 \mathbf{u}_{2,0} + \alpha\beta \mathbf{u}_{1,1} 
	+ \beta^2 \mathbf{u}_{0,2}  
	+ \dots,\label{eq:u_initial_pert_exp_axisym}\\
	p &= p_{0,0} + \alpha p_{1,0} + \beta p_{0,1} 
	+ \alpha^2 p_{2,0} + \alpha\beta p_{1,1} 
	+ \beta^2 p_{0,2} 
	+ \dots,\label{eq:p_initial_pert_exp_axisym}
\end{align}
i.e.~the velocity and pressure at order $\alpha^{m}\beta^{n}$ are given by $\mathbf{u}_{m,n}$ and $p_{m,n}$, respectively.
We note that had we chosen to model the bulk viscosity as depending linearly on temperature, we would have posed above a perturbation expansion in three small parameters instead of two, with the third parameter being a thermal bulk viscosity coefficient.
In the following sections, we will solve for the flow at each order.

\subsection{Solution at order $\beta^{n}$}\label{sec:beta_n2_gamma_n3}

First, as was shown for the parallel-plate geometry~\cite{liao2023theoretical}, we claim that in Eq.~\eqref{eq:u_initial_pert_exp_axisym} and Eq.~\eqref{eq:p_initial_pert_exp_axisym}, we may set both the velocity and pressure to be zero at orders $\beta^{n}$ for all $n$, with the physical interpretation that thermal expansion is essential for the fluid flow. 
To see this, we observe that if we set the thermal expansion coefficient $\alpha$ to be zero, then the terms at order $\alpha^{m}\beta^{n}$ with $m \geq 1$ in Eq.~\eqref{eq:u_initial_pert_exp_axisym} and Eq.~\eqref{eq:p_initial_pert_exp_axisym} vanish, while the mass conservation equation [Eq.~\eqref{eq:mass_consn_3D_axisym_dimless}] becomes
\begin{align}
	\nabla \cdot \mathbf{u} &= 0.\label{eq:mass_beta_n}
\end{align}
We observe that zero flow, $\mathbf{u}=\mathbf{0}$ and $p=0$, solves Eq.~\eqref{eq:mass_beta_n} together with Eq.~\eqref{eq:momentum}. 
By expanding this solution in the parameter $\beta$, we then see that this corresponds to $\mathbf{u}_{0,n}=\mathbf{0}$ and $p_{0,n}=0$, as claimed. 
The perturbation expansions then simplify to
\begin{align}
	\mathbf{u} &=  \alpha \mathbf{u}_{1,0} 
	+ \alpha^2 \mathbf{u}_{2,0} + \alpha\beta \mathbf{u}_{1,1} 
	+ \dots,\label{eq:u_pert_exp_axisym}\\
	p &= \alpha p_{1,0}
	+ \alpha^2 p_{2,0} + \alpha\beta p_{1,1} 
	+ \dots.\label{eq:p_pert_exp_axisym}
\end{align}
The structure of these expansions reflects the physics of the flows.
Every term includes the thermal expansion coefficient $\alpha$, which mathematically captures the fact that thermal expansion drives the flow, via forcing in the mass conservation equation.
This flow is an automatic consequence of the spatio-temporally varying temperature field~\cite{yariv2004flow}; we emphasise that it is independent of gravity, thus distinguishing it from buoyancy-driven flows.

We are now in a position to revisit the assumption to neglect advection of heat made in Sec.~\ref{sec:temperature} for the thermal problem.
Since the instantaneous flow occurs at order~$\alpha$, the ratio of the advection term  $\mathbf{u}\cdot\nabla T$ to the rate of change of temperature at a fixed position due to scanning $\partial T/\partial t$ is an order-$\alpha$ quantity; thus, we confirm that we may neglect advection of heat by fluid flow in favour of the unsteady term in the leading-order thermal problem.

\subsection{Velocity field associated with the time-variation of the heat-spot amplitude}\label{sec:switching_on_flow}

We may decompose the fluid flow induced by the scanning heat spot into two contributions, in a precise manner that we explain mathematically below: one associated with the time variation of the heat-spot amplitude, and the other related to the translation of the heat spot. 
In this section, we consider the former and derive the contribution to the velocity field that captures the switching-on and switching-off of a spherically symmetric heat spot.

\subsubsection{Decomposition of velocity field}\label{sec:decomposition}

First, we introduce spherical polar coordinates $(R,\theta,\phi)$ with origin at the centre of the translating heat spot, i.e.~at $(r=0,z=t)$, so that the spherical radial coordinate is given by 
	\begin{align}
		R=\sqrt{r^2+(z-t)^2}.\label{eq:spherical_R}
	\end{align} 
We assume that the temperature perturbation during one scan of the heat spot has the form
\begin{align}
	\Delta T(R,t) = A(t) \Theta(R)\label{eq:temp_general_axisym},
\end{align}
i.e.~an arbitrary, time-dependent amplitude $A(t)$, which is zero at the ends of the scan path, multiplied by a shape function $\Theta(R)$ that translates in the $z$ direction and decays at infinity. 
The spherical symmetry of the (instantaneous) heat-spot shape is mathematically convenient for this first model in three dimensions.
In this case, using Eqs.~\eqref{eq:density_temp_dimless_axisym},~\eqref{eq:spherical_R} and \eqref{eq:temp_general_axisym}, the mass conservation equation [Eq.~\eqref{eq:mass_consn_3D_axisym_dimless}] becomes
\begin{align}
	-\alpha \left (A'(t)\Theta(R) + A(t)\Theta'(R) \frac{\partial R}{\partial t}\right ) + \nabla \cdot(\rho \mathbf{u}) = 0.\label{eq:mass_consn}
\end{align}
The two forcing terms reflect the two reasons that the temperature field at a given position varies with time: because the amplitude is time-dependent (first term) and because the shape of the temperature perturbation translates in space (second term).
Since Eq.~\eqref{eq:mass_consn} is linear in velocity, we may decompose the solution into two contributions (as in Ref.~\cite{liao2023theoretical}), to account for the two forcing terms.
We thus write the velocity field as
\begin{align}
	\mathbf{u} = \mathbf{u}^\text{(S)} + \mathbf{u}^\text{(T)}.\label{eq:decomposition_u_S_T}
\end{align}
Here, we introduce the switching-on velocity field $\mathbf{u}^\text{(S)}$, associated with the time-variation of the heat-spot amplitude, as a solution to the equation given by 
\begin{align}
	- \alpha A'(t) \Theta(R) + \nabla \cdot (\rho \mathbf{u}^\text{(S)}) = 0,\label{eq:switch_mass_consn}
\end{align}
i.e.~treating the forcing term containing $A'(t)$, while the velocity field $\mathbf{u}^\text{(T)}$ associated with translation of the heat spot satisfies
\begin{align}
	- \alpha A(t)\Theta'(R) \frac{\partial R}{\partial t} + \nabla \cdot (\rho \mathbf{u}^\text{(T)}) = 0,
\end{align}
i.e.~accounting for the forcing term containing $\Theta'(R) \frac{\partial R}{\partial t}$.

\subsubsection{Mathematical derivation of switching-on flow}\label{sec:switch_derive}

We treat in this section the switching-on velocity field, associated with the time-variation of the heat-spot amplitude; we will return to the translational contribution in Sec.~\ref{sec:alpha}.
The switching-on flow $\mathbf{u}^\text{(S)}$ satisfies Eq.~\eqref{eq:switch_mass_consn}, a version of the mass conservation equation but forced by a regularised source, instantaneously centred at the location of the heat spot and with time-varying amplitude.
We assume that this switching-on velocity field also satisfies the momentum equation with a corresponding switching-on pressure field, is non-singular at the origin  and decays at infinity (provided the temperature perturbation decays sufficiently fast at infinity).

We now pose a spherically symmetric ansatz for the switching-on velocity field, given by
\begin{align}
	\mathbf{u}^\text{(S)} = u^\text{(S)}(R,t)\mathbf{e}_R,\label{eq:switch_ansatz}
\end{align}
where $\mathbf{e}_R$ is the radial unit vector from the centre of the heat spot.
The mass conservation equation [Eq.~\eqref{eq:switch_mass_consn}] in spherical polar coordinates then becomes
\begin{align}
	-\alpha A'(t) \Theta(R) + \frac{1}{R^2} \frac{\partial }{\partial R}(R^2 \rho u^\text{(S)}) = 0.\label{eq:mass_consn_switch}
\end{align}

\subsubsection{Flow result and physical interpretation}
	
Integrating the defining mass conservation equation for $\mathbf{u}^\text{(S)}$ [Eq.~\eqref{eq:mass_consn_switch}] and applying the boundary conditions (Sec.~\ref{sec:switch_derive}), we obtain the switching-on flow~[Eq.~\eqref{eq:switch_ansatz}], for an arbitrary, spherically symmetric temperature perturbation of the form in Eq.~\eqref{eq:temp_general_axisym}, as
\begin{align}
	u^\text{(S)}(R,t) = \frac{\alpha A'(t)}{\rho(R,t) R^2} \int_0^R \tilde{R}^2 \Theta(\tilde{R}) \, d\tilde{R}.\label{eq:u_switch}
\end{align}
We observe that the perturbation expansion for this switching-on velocity $\mathbf{u}^\text{(S)}$ features only powers of the thermal expansion coefficient~$\alpha$ (from Taylor expanding the factor of $\rho^{-1}$), but not the other dimensionless parameter $\beta$, which characterises thermal shear viscosity changes. 
It is also independent of the bulk viscosity~$\kappa$.
As in two dimensions~\cite{liao2023theoretical}, this switching-on contribution can be derived from only mass conservation and spherical symmetry of the prescribed temperature perturbation, independent of the momentum equation and hence independent of viscosity; it is a kinematic phenomenon that relies solely on thermal expansion. 
The flow~$\mathbf{u}^\text{(S)}$ also fully accounts for any appearances of the rate of change of heat-spot amplitude, $A'(t)$, in the solution for the full flow $\mathbf{u}$ due to the spherically symmetric heat spot. 
Physically, when the heat-spot amplitude is increasing [$A'(t)>0$], the switching-on flow is an instantaneous regularised source, as the fluid expands radially outwards. 
Conversely, when the heat-spot amplitude is decreasing [$A'(t)<0$], the switching-on flow is instead an instantaneous sink, reflecting the contraction of the fluid as temperature decreases locally.

\subsection{Solution at order $\alpha$}\label{sec:alpha}

We now proceed to solve in this section for the leading-order instantaneous flow during a scan, which occurs at order $\alpha$ and is purely driven by thermal expansion. 
We will show that this consists of a contribution due to the switching-on of the heat spot and another related to the translation of the heat spot.

\subsubsection{General solution}

With Eqs.~\eqref{eq:density_temp_dimless_axisym} and~\eqref{eq:u_pert_exp_axisym} for the density and velocity, respectively, we expand Eq.~\eqref{eq:mass_consn_3D_axisym_dimless} to obtain that at order~$\alpha$, the statement of mass conservation is given by
\begin{align}
	-\frac{\partial \Delta T}{\partial t} + \nabla \cdot \mathbf{u}_{1,0} = 0,\label{eq:mass_consn_alpha}
\end{align}
while the momentum equation [Eq.~\eqref{eq:momentum}] is given by
\begin{align}
	- \nabla p_{1,0}  +  \nabla \left [ \left ( \kappa +  \frac{1}{3} \right ) \nabla \cdot \mathbf{u}_{1,0} \right ] + \nabla^2 \mathbf{u}_{1,0} = \mathbf{0}.\label{eq:momentum_alpha}
\end{align}
Building on the decomposition of the flow [Eq.~\eqref{eq:decomposition_u_S_T}] introduced in Sec.~\ref{sec:decomposition}, we pose an ansatz for the velocity field at order~$\alpha$ given by
\begin{align}
	\mathbf{u}_{1,0} &= \frac{\partial }{\partial t} [A(t) \mathbf{u}_{1,0}^\text{(S)}] \nonumber\\
	& \equiv A'(t)\mathbf{u}_{1,0}^\text{(S)} + A(t) \mathbf{u}_{1,0}^\text{(T)},\label{eq:u100_general_decomposn}
\end{align}
for $-t_0 \leq t \leq t_0$, where $\mathbf{u}_{1,0}^\text{(S)}$ is the velocity field associated with the time-variation of the heat-spot amplitude and $\mathbf{u}_{1,0}^\text{(T)}$ is the velocity field associated with the translation of the heat spot. 
Its structure is inherited from the forcing term in the mass conservation equation [Eq.~\eqref{eq:mass_consn_alpha}], which takes the form of a time-derivative of a function proportional to the amplitude $A(t)$. 
Here, we define the switching-on velocity field $\mathbf{u}_{1,0}^\text{(S)}$ at order $\alpha$ via a perturbation expansion of the switching-on velocity field $\mathbf{u}^\text{(S)}$ [Eq.~\eqref{eq:u_switch}] derived in Sec.~\ref{sec:switching_on_flow}, given by
\begin{align}
	\mathbf{u}^\text{(S)} &= \left (\frac{\alpha A'(t)}{R^2} \int_0^R \tilde{R}^2 \Theta(\tilde{R}) \, d\tilde{R} + O(\alpha^2) \right )\mathbf{e}_R \nonumber\\
	&\equiv \alpha A'(t) \mathbf{u}_{1,0}^\text{(S)} + O(\alpha^2).\label{eq:u10S_uS_expansion}
\end{align}
It may be verified that the ansatz for the flow at order $\alpha$ satisfies the mass conservation and momentum equations, with the pressure field [found by taking the divergence of the momentum equation, i.e.~Eq.~\eqref{eq:momentum_alpha}, and combining with mass conservation] given by
\begin{align}
	p_{1,0} = \left ( \kappa  + \frac{4}{3}\right ) \frac{\partial \Delta T}{\partial t}.
\end{align}
We emphasise that we did not need to assume that the bulk viscosity $\kappa$ is a constant; instead, it depends on space and time via the temperature field.
While the pressure depends on bulk viscosity, we note that the velocity field at order $\alpha$ is independent of bulk viscosity.
We also note that the velocity field at order $\alpha$ during a scan is the time-derivative of a function proportional to the heat-spot amplitude.
Consequently, when integrated over a scan, this will give rise to zero net displacement of material points at order $\alpha$, just as for the parallel-plate setup in Ref.~\cite{liao2023theoretical}.

\subsubsection{Flow result for Gaussian temperature profile and physical interpretation}

\begin{figure}[t]
		\centering
	{\includegraphics[width=0.8\textwidth]{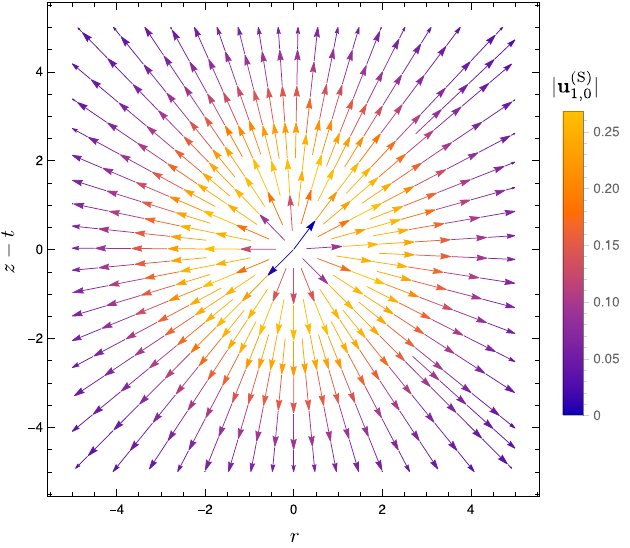}
		\caption{Streamlines of the flow $\mathbf{u}_{1,0}^\text{(S)}$ at order $\alpha$ associated with the switching-on of the spherical heat spot, with centre at $(r=0,z=t)$ in dimensionless coordinates (nondimensionalisation described in Sec.~\ref{sec:nondiml}). Colour shows the magnitude $\vert \mathbf{u}_{1,0}^\text{(S)} \vert $.
		}
		\label{fig:plot_u100S_axisym_FLUCS_06022024}}
\end{figure}

To illustrate our solution for the flow at order $\alpha$ during a scan given by Eqs.~\eqref{eq:u100_general_decomposn}--\eqref{eq:u10S_uS_expansion}, we now prescribe a Gaussian temperature profile for the spherical heat spot, motivated in Sec.~\ref{sec:temperature} and given by
\begin{align}
	\Delta T(R,t) = A(t)\exp(-R^2/2).\label{eq:Gaussian_temp_dimless}
\end{align}
For this Gaussian temperature profile, the switching-on velocity field at order $\alpha$, from Eq.~\eqref{eq:u10S_uS_expansion}, is given by
\begin{align}
	\mathbf{u}_{1,0}^\text{(S)} &= \frac{1}{R^2} \int_0^R \tilde{R}^2 \exp(-\tilde{R}^2/2) \, d\tilde{R} \, \mathbf{e}_R\nonumber\\
	&\equiv \left (
	\frac{\sqrt{\pi}\erf(R/\sqrt{2})}{\sqrt{2}R^2}
	-\frac{\exp(-R^2/2)}{R}
	\right ) \mathbf{e}_R.\label{eq:u10S}
\end{align}
This purely radial flow, shown in Fig.~\ref{fig:plot_u100S_axisym_FLUCS_06022024}, is a hydrodynamic source in the far field, decaying as $1/R^2$ (since the fluid is three-dimensional and unbounded).
As for the parallel-plate case in Ref.~\cite{liao2023theoretical}, this arises from local volume increase of the fluid due to heating. 

\begin{figure}[t]
		\centering
	{\includegraphics[width=0.8\textwidth]{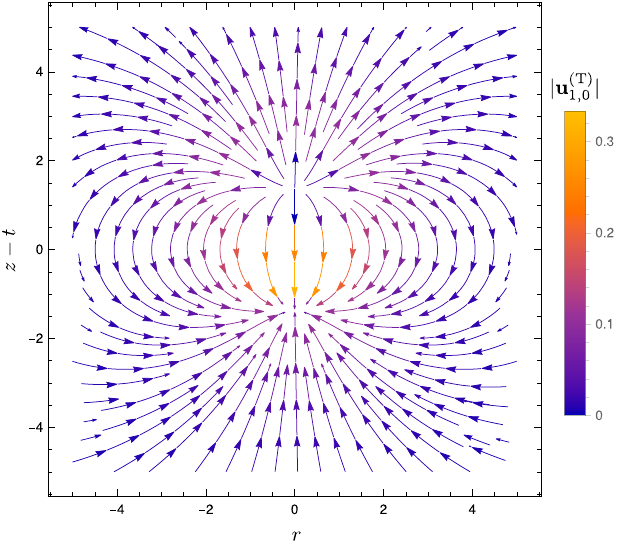}
		\caption{Streamlines of the instantaneous flow $\mathbf{u}_{1,0}^\text{(T)}$ at order $\alpha$ associated with the translation of the spherical heat spot, during one scan. The flow is axisymmetric about the $z$ axis. Colour indicates the magnitude $\vert \mathbf{u}_{1,0}^\text{(T)} \vert$.
		}
		\label{fig:plot_u100T_axisym_FLUCS_06022024}}
\end{figure}

The velocity field associated with the translation of the heat spot, from Eqs.~\eqref{eq:u100_general_decomposn} and~\eqref{eq:u10S}, and shown in Fig.~\ref{fig:plot_u100T_axisym_FLUCS_06022024}, is given by
\begin{align}
	\mathbf{u}_{1,0}^\text{(T)} \equiv& \frac{\partial}{\partial t} \mathbf{u}_{1,0}^\text{(S)} \nonumber\\
	=& \cos\theta\left (
	\frac{\sqrt{2\pi}\erf(R/\sqrt{2})}{R^3}
	- \left ( 1+\frac{2}{R^2}\right )\exp(-R^2/2) 
	\right )\mathbf{e}_R \nonumber\\
	&+ \sin\theta \left (\frac{\sqrt{\pi}\erf(R/\sqrt{2})}{\sqrt{2}R^3}
	- \frac{\exp(-R^2/2)}{R^2}\right ) \mathbf{e}_\theta,\label{eq:u10T}
\end{align}
where $\mathbf{e}_\theta$ is the spherical basis vector corresponding to the polar angle $\theta$. 
The streamlines are qualitatively the same as those for the parallel-plate setup~\cite{liao2023theoretical}. 
In the far field, the flow is a source dipole in the far field, decaying as $1/R^3$.
The source at the front corresponds to the arrival of the heat spot, while the sink at the back results from cooling due to the departure of the heat spot, just as for the parallel-plate geometry~\cite{weinert2008optically,liao2023theoretical}. 

In terms of dimensional variables, the leading-order flow during one scan scales as~$\alpha\Delta T_0 U$; it is proportional to the thermal expansion coefficient, the peak temperature change and the velocity of translation of the heat spot. 
Hence, the leading-order instantaneous flow scales linearly with the scan frequency. 
Furthermore, changing the sign of the thermal expansion coefficient reverses this flow; for example, water has a positive value of $\alpha$ at room temperature, but negative when below~$4~\si{\degreeCelsius}$. 

Note that for all plots of instantaneous fluid flow during one scan in Sec.~\ref{sec:flow}, the vertical axis label is $z-t$, reflecting that the origin of our spherical polar coordinate system is at the centre of the heat spot~$(r=0,z=t)$.
Thus, the plots show the flow as viewed when moving at the same velocity as the heat spot.

\subsection{Solution at order $\alpha\beta$}

For water and glycerol (common liquids used in experiments), the thermal expansion coefficient~$\alpha$ is much smaller than the thermal shear viscosity coefficient~$\beta$, so we may expect effects at order $\alpha\beta$ to be more visible in experiments using these fluids than those at order $\alpha^2$, as was the case for the parallel-plate setup~\cite{liao2023theoretical}. 
We therefore now consider the solution for instantaneous thermoviscous flow during one scan at order~$\alpha\beta$, and will next return to the purely thermal expansion-driven effect at order~$\alpha^2$.

\subsubsection{Mathematical derivation}

At order $\alpha\beta$, mass conservation [Eq.~\eqref{eq:mass_consn_3D_axisym_dimless}] is given by
\begin{align}
	\nabla \cdot \mathbf{u}_{1,1} = 0,\label{eq:mass_consn_alphabeta}
\end{align}
i.e.~the flow at order $\alpha\beta$ is incompressible. Here we have substituted Eqs.~\eqref{eq:density_temp_dimless_axisym} (density) and~\eqref{eq:u_pert_exp_axisym} (velocity perturbation expansion), and used the fact that there is no flow at order $\beta$ (Sec.~\ref{sec:beta_n2_gamma_n3}).

Expanding Eq.~\eqref{eq:momentum}, the momentum equation at order $\alpha\beta$ is given by
\begin{align}
	&- \nabla p_{1,1} 
	+ \nabla (\kappa \nabla \cdot \mathbf{u}_{1,1})
	+ \nabla^2 \mathbf{u}_{1,1} 
	- \Delta T \nabla^2 \mathbf{u}_{1,0}
	+ \frac{1}{3}  \nabla (\nabla \cdot \mathbf{u}_{1,1})\nonumber\\
	&- \frac{1}{3} \Delta T \nabla (\nabla \cdot \mathbf{u}_{1,0}) 
	- (\nabla (\Delta T)) \cdot [\nabla \mathbf{u}_{1,0} + (\nabla \mathbf{u}_{1,0})^\text{T}]
	+ \frac{2}{3} (\nabla (\Delta T)) (\nabla \cdot \mathbf{u}_{1,0}) \nonumber\\
	&= \mathbf{0}.
\end{align} 
Using mass conservation at orders $\alpha$ [Eq.~\eqref{eq:mass_consn_alpha}] and $\alpha\beta$ [Eq.~\eqref{eq:mass_consn_alphabeta}] to simplify this, we find
\begin{align}
	- \nabla p_{1,1} 
	+ \nabla^2 \mathbf{u}_{1,1} 
	- \Delta T \nabla^2 \mathbf{u}_{1,0}
	- \frac{1}{3} \Delta T \nabla \frac{\partial\Delta T }{\partial t} 
	- (\nabla (\Delta T)) \cdot [\nabla \mathbf{u}_{1,0} + (\nabla \mathbf{u}_{1,0})^\text{T}] & \nonumber\\
	+ \frac{2}{3} \nabla (\Delta T)  \frac{\partial \Delta T }{\partial t}
	&= \mathbf{0}.\label{eq:momentum_alphabeta}
\end{align}
We observe that due to incompressibility of the flow at order $\alpha\beta$ [Eq.~\eqref{eq:mass_consn_alphabeta}], the bulk viscosity $\kappa$ does not feature in this equation, which takes the form of the forced Stokes equation; the solution for the velocity field $\mathbf{u}_{1,1}$ will correspondingly also be independent of bulk viscosity. 

We now use a streamfunction approach to solve for the flow.
The velocity field at order $\alpha\beta$ is divergence-free, so we may write it in terms of a Stokes streamfunction $\Psi$ as
\begin{align}
	\mathbf{u}_{1,1} = \nabla \times \left ( -\frac{\Psi(R,\theta,t)}{R\sin\theta} \mathbf{e}_\phi \right ),\label{eq:u11_streamfunction_curl}
\end{align}
where $\mathbf{e}_\phi$ is the spherical basis vector corresponding to the azimuthal angle $\phi$.
It is a classical result~\cite{happel} that
\begin{align}
	\nabla^2 (\nabla \times \mathbf{u}_{1,1}) = \frac{\E^2(\E^2 \Psi)}{R\sin\theta} \mathbf{e}_\phi,
\end{align}
where the operator $\E^2$ is given by
\begin{align}
	\E^2 \Psi \equiv \frac{\partial^2 \Psi}{\partial R^2} + \frac{\sin\theta}{R^2} \frac{\partial}{\partial \theta } \left (\frac{1}{\sin\theta} \frac{\partial\Psi}{\partial\theta} \right ).\label{eq:E2_operator}
\end{align}
Using this, the $\phi$ component of the curl of the momentum equation [Eq.~\eqref{eq:momentum_alphabeta}] in spherical polar coordinates becomes
\begin{align}
	\frac{\E^2(\E^2 \Psi)}{R\sin\theta} = A(t)^2\sin\theta \left (4R\exp(-R^2) + \frac{6}{R}\exp(-R^2) - \frac{3\sqrt{2\pi}}{R^2}\exp(-R^2/2)\erf(R/\sqrt{2}) \right) .
\end{align}
Again, as with the parallel-plate setup~\cite{liao2023theoretical}, we observe that the forcing at order $\alpha\beta$ does not contain $A'(t)$; only the flow fields at orders $\alpha^n$ (solely driven by thermal expansion) depend on the rate of change of the heat-spot amplitude, through the switching-on contribution.

To solve this, we choose the ansatz
\begin{align}
	\Psi = A(t)^2 f(R) \sin^2\theta,
\end{align}
which is similar to that for the incompressible Stokes flow past a rigid sphere~\cite{happel}.
By reducing the problem to two second-order ordinary differential equations in $R$ (one for each application of the operator $\E^2$), this allows us to solve for the axisymmetric streamfunction $\Psi$ as
\begin{align}
	\Psi 
	=  A(t)^2 \sin^2\theta \bigg \{& \frac{\sqrt{\pi}}{20 R} \erf(R)    
	+ \left (\frac{\sqrt{\pi}}{5\sqrt{2}}R^3 + 2\sqrt{2\pi} R  - \frac{\sqrt{2\pi}}{5R} \right ) \exp(-R^2/2) \erf(R/\sqrt{2})  \nonumber\\
	&+ \left (\frac{\pi}{20} R^4  + \frac{\pi}{4} R^2 \right )  [\erf(R/\sqrt{2})^2 - 1 ] + \left (\frac{1}{10}R^2 + \frac{3}{10 }\right ) \exp(-R^2) \bigg\}.
\end{align}
Applying Eq.~\eqref{eq:u11_streamfunction_curl} then gives the corresponding velocity field at order $\alpha\beta$, which satisfies the boundary conditions at the origin and at infinity.

\subsubsection{Flow result and physical interpretation}\label{sec:u11_physical_interpret}

\begin{figure}[t]
		\centering
	{\includegraphics[width=0.8\textwidth]{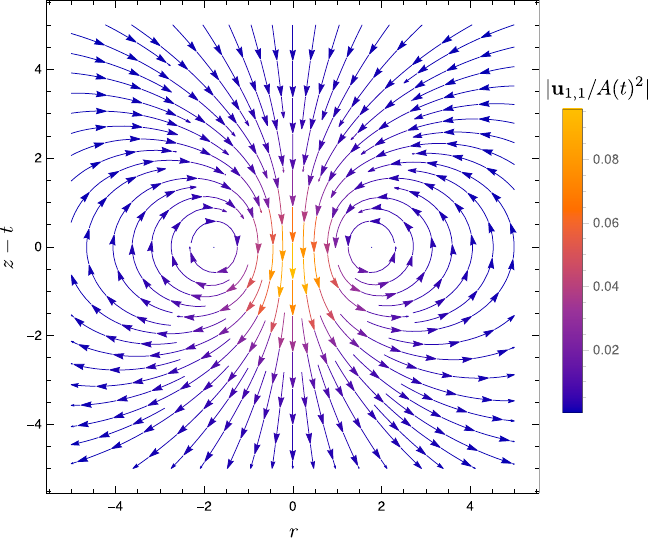}
		\caption{Streamlines of the flow $\mathbf{u}_{1,1}$ at order $\alpha\beta$ due to a spherical heat spot, translating in the $z$ direction. Colour shows the magnitude of the scaled flow $\vert \mathbf{u}_{1,1} / A(t)^2 \vert $.
		}
		\label{fig:plot_u110T_axisym_FLUCS_02022024}}
\end{figure}

The velocity field $\mathbf{u}_{1,1}$ at order $\alpha\beta$, derived above, is given in spherical polar coordinates by
\begin{align}
	\mathbf{u}_{1,1} = A(t)^2 \Bigg \{& \mathbf{e}_R\cos\theta \bigg \{ - \frac{\sqrt{\pi}}{10R^3} \erf(R)  \nonumber\\
	&+ \left (- \frac{\sqrt{2\pi}}{5} R - \frac{4\sqrt{2\pi}}{5R} + \frac{2\sqrt{2\pi}}{5R^3}\right ) \exp(-R^2/2) \erf(R/\sqrt{2})   \nonumber\\
	&+ \left ( \frac{\pi}{10}R^2 + \frac{\pi}{2} \right ) [ 1- \erf(R/\sqrt{2})^2 ] - \left ( \frac{1}{5} + \frac{3}{5R^2}  \right ) \exp(-R^2) 
	\bigg \} \nonumber\\
	&+ \mathbf{e}_\theta \sin\theta \bigg \{ 
	- \frac{\sqrt{\pi}}{20R^3} \erf(R)  \nonumber\\
	&+ \left (\frac{2\sqrt{2\pi}}{5} R  + \frac{3\sqrt{2\pi}}{5R} + \frac{\sqrt{2\pi}}{5R^3} \right ) \exp(-R^2/2) \erf(R/\sqrt{2})  \nonumber\\
	&  + \left ( \frac{\pi}{5} R^2 + \frac{\pi}{2} \right ) [\erf(R/\sqrt{2})^2 - 1 ]  +  \left (\frac{2}{5}  - \frac{3}{10R^2} \right ) \exp(-R^2) 
	\bigg \} \Bigg \},\label{eq:u11}
\end{align}
for $-t_0 \leq t \leq t_0$.
We plot the streamlines of this flow in Fig.~\ref{fig:plot_u110T_axisym_FLUCS_02022024} and note that it qualitatively matches that for the parallel-plate setup at the same order~\cite{liao2023theoretical}.

In the far field, the flow at order $\alpha\beta$ is given by
\begin{align}
	\mathbf{u}_{1,1} \sim - \frac{A(t)^2 \sqrt{\pi}}{20R^3} ( 2  \cos\theta \mathbf{e}_R+  \sin\theta \mathbf{e}_\theta).\label{eq:u11_far}
\end{align}
This is a three-dimensional source dipole, decaying as $1/R^3$, with a sink at the front and a source at the back, the same physics as for the parallel-plate geometry~\cite{liao2023theoretical}. 
Strikingly, the flow on the $z$ axis is downwards, i.e.~in the opposite direction to heat-spot translation. 
To explain the flow $\mathbf{u}_{1,1}$ at order $\alpha\beta$ physically, we consider it as a modification, due to thermal shear viscosity changes, of the flow contribution $A(t) \mathbf{u}_{1,0}^\text{(T)}$ at order $\alpha$ associated with translation of the heat spot.
Specifically, we may view the near field of $\mathbf{u}_{1,1}$ as localised amplification of the front-to-back flow near the heat spot of $A(t) \mathbf{u}_{1,0}^\text{(T)}$, due to reduced shear viscosity locally, while the far-field source dipole of $\mathbf{u}_{1,1}$ enforces incompressibility at this order.

Dimensionally, the thermoviscous flow at order $\alpha\beta$ during one scan scales with $\alpha\beta\Delta T_0^2 U$.
Like the leading-order flow at order $\alpha$, the order-$\alpha\beta$ contribution is proportional to the velocity of the heat spot, and would be reversed by a change in sign of the thermal expansion coefficient.
However, in contrast with the leading order, the flow at order $\alpha\beta$ scales quadratically with peak temperature change.

We will see in Sec.~\ref{sec:transport} that this instantaneous thermoviscous flow at order $\alpha\beta$ during a scan will contribute to the leading-order time-averaged velocity of tracers. 

\subsection{Solution at order $\alpha^2$}\label{sec:alpha_sq}

To complete the analysis of the instantaneous flow correct to quadratic order, we now consider the flow at order $\alpha^2$.
This flow does not depend on thermal shear viscosity changes, instead relying only on thermal expansion; it is thus fundamentally different to the order-$\alpha\beta$ thermoviscous flow contribution and would exist even if the shear viscosity were constant with respect to temperature.

\subsubsection{Mathematical derivation}

Expanding Eq.~\eqref{eq:mass_consn_3D_axisym_dimless} [again using Eqs.~\eqref{eq:density_temp_dimless_axisym} and~\eqref{eq:u_pert_exp_axisym}], we find that the mass conservation equation at order $\alpha^2$ is given by 
\begin{align}
	\nabla\cdot\mathbf{u}_{2,0} - \nabla\cdot(\Delta T\mathbf{u}_{1,0})=0.\label{eq:mass_consn_alpha_sq}
\end{align}
In contrast with order $\alpha\beta$, the flow at order $\alpha^2$ is not incompressible.
However, to solve for the flow, it is useful to define an incompressible modified velocity field $\mathbf{v}$ as
\begin{align}
	\mathbf{v}\equiv \mathbf{u}_{2,0}-\Delta T\mathbf{u}_{1,0}.\label{eq:modified_vel_v}
\end{align}
From Eq.~\eqref{eq:momentum}, at order $\alpha^2$, the momentum equation is given by
\begin{align}
	- \nabla p_{2,0}
	+ \nabla ( \kappa \nabla \cdot \mathbf{u}_{2,0})
	+  \nabla^2 \mathbf{u}_{2,0} 
	+ \frac{1}{3}  \nabla (\nabla \cdot \mathbf{u}_{2,0})
	= \mathbf{0}.
\end{align}
Using mass conservation [Eq.~\eqref{eq:mass_consn_alpha_sq}], we simplify this to
\begin{align}
	- \nabla p_{2,0}
	+ \nabla \left [\left (\kappa + \frac{1}{3}\right ) \nabla\cdot(\Delta T\mathbf{u}_{1,0}) \right ]
	+  \nabla^2 \mathbf{u}_{2,0} 
	= \mathbf{0}.
\end{align}
Taking the curl, we obtain
\begin{align}
	\nabla^2 (\nabla \times \mathbf{u}_{2,0} )= \mathbf{0},
\end{align}
thus eliminating the gradient terms. In particular, the bulk viscosity term vanishes again, so that the solution for the flow $\mathbf{u}_{2,0}$ at order $\alpha^2$ will be independent of bulk viscosity. 
Rewriting this in terms of the modified velocity field [Eq.~\eqref{eq:modified_vel_v}] gives
\begin{align}
	\nabla^2 [\nabla \times (\mathbf{v} + \Delta T\mathbf{u}_{1,0})] = \mathbf{0}.\label{eq:curl_momentum_v}
\end{align}
The solution method now closely follows that at order $\alpha\beta$: since the modified velocity field is divergence-free, we may write it in terms of a Stokes streamfunction $\Phi$ as
\begin{align}
	\mathbf{v} = \nabla \times \left ( -\frac{\Phi(R,\theta,t)}{R\sin\theta} \mathbf{e}_\phi \right ).\label{eq:modified_vel_v_Stokes_stream_function}
\end{align}
Similarly to order $\alpha\beta$, we obtain the relation
\begin{align}
	\nabla^2 (\nabla \times \mathbf{v}) = \frac{\E^2(\E^2 \Phi)}{R\sin\theta} \mathbf{e}_\phi,
\end{align}
where the operator $\E^2$ is given by Eq.~\eqref{eq:E2_operator}.
We then write Eq.~\eqref{eq:curl_momentum_v} in terms of the streamfunction as 
\begin{align}
	\frac{\E^2(\E^2 \Phi)}{R\sin\theta} \mathbf{e}_\phi = -  \nabla^2 [\nabla \times ( \Delta T\mathbf{u}_{1,0})].\label{eq:stream_fn_eq_alpha_sq}
\end{align}
The right-hand side of Eq.~\eqref{eq:stream_fn_eq_alpha_sq} is given explicitly by
\begin{align}
	-  \nabla^2 [\nabla \times ( \Delta T\mathbf{u}_{1,0}) ]
	= A(t)^2 \sin\theta
	\Bigg [&\frac{\sqrt{\pi}}{\sqrt{2}} \left (1 + \frac{1}{R^2} \right )\exp(-R^2/2) \erf(R/\sqrt{2}) \nonumber\\
	&- \left (4R + \frac{1}{R} \right ) \exp(-R^2/2)
	\Bigg ] \mathbf{e}_\phi.
\end{align}
This does not contain the rate of change of the heat-spot amplitude $A'(t)$, since the contribution associated with the switching-on of the heat spot involves the curl of a function of $R$ parallel to the radial direction ($ \Delta T\mathbf{u}_{1,0}^\text{(S)}$); this curl is zero by symmetry. 
As a result, we will see that the modified velocity field $\mathbf{v}$ does not contain $A'(t)$. Any contributions to the velocity field $\mathbf{u}_{2,0}$ at order $\alpha^2$ involving the rate of change of heat-spot amplitude therefore originate from the term $\Delta T \mathbf{u}_{1,0}$ in Eq.~\eqref{eq:modified_vel_v}, specifically from~$\Delta T \mathbf{u}_{1,0}^\text{(S)}$.

Now, the streamfunction equation [Eq.~\eqref{eq:stream_fn_eq_alpha_sq}] becomes
\begin{align}
	\E^2(\E^2 \Phi)
	&= 
	A(t)^2 \sin^2\theta\left [
	\frac{\sqrt{\pi}}{\sqrt{2}} \left (R + \frac{1}{R} \right )\exp(-R^2/2) \erf(R/\sqrt{2}) 
	- \left (4R^2 + 1 \right ) \exp(-R^2/2)
	\right ].
\end{align}
Following the same method as at order $\alpha\beta$, we choose the ansatz
\begin{align}
	\Phi = A(t)^2 g(R) \sin^2\theta,\label{eq:Phi_ansatz}
\end{align}
and find that the function $g(R)$ is given by
\begin{align}
	g(R)= \frac{1}{12} \Bigg(&-\frac{\sqrt{\pi } \erf(R)}{R} -2 \sqrt{2 \pi } R  \exp(-R^2/2) \erf(R/\sqrt{2})+\frac{2 \sqrt{2 \pi } \exp(-R^2/2) \erf(R/\sqrt{2})}{R}\nonumber\\
	&-2 \exp(-R^2) + \pi  R^2-\pi  R^2 \erf(R/\sqrt{2})^2\Bigg).\label{eq:Phi_ansatz_g_result}
\end{align}
The Stokes streamfunction for the modified velocity field is therefore given by Eqs.~\eqref{eq:Phi_ansatz}--\eqref{eq:Phi_ansatz_g_result}. We can substitute this into Eq.~\eqref{eq:modified_vel_v_Stokes_stream_function} to find the modified velocity field, and hence obtain the flow $\mathbf{u}_{2,0}$ at order $\alpha^2$ from Eq.~\eqref{eq:modified_vel_v}.

\subsubsection{Flow result and physical interpretation}\label{sec:u20_result}

\begin{figure}[t]
	\centering
	{\includegraphics[width=0.8\textwidth]{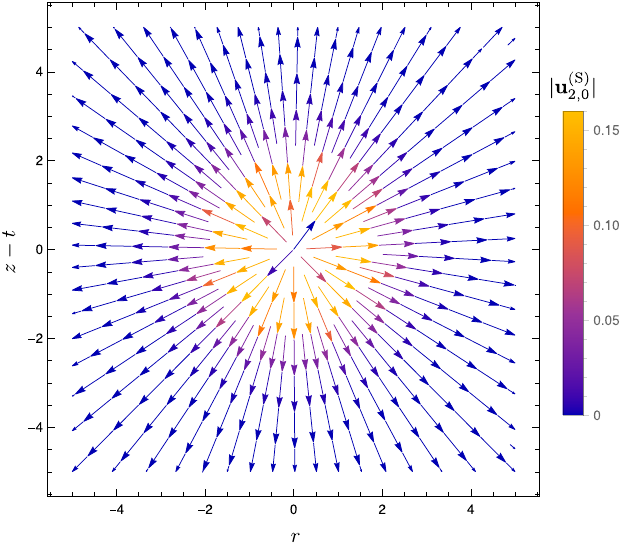}
		\caption{Streamlines of the radial flow $\mathbf{u}_{2,0}^\text{(S)}$ at order $\alpha^2$ associated with the switching-on of a spherical heat spot, with centre at $(r=0,z=t)$. Colour shows the magnitude $\vert \mathbf{u}_{2,0}^\text{(S)} \vert $.
		}
		\label{fig:plot_u200S_axisym_FLUCS_06022024}}
\end{figure}
\begin{figure}[t]
	\centering
	{\includegraphics[width=0.8\textwidth]{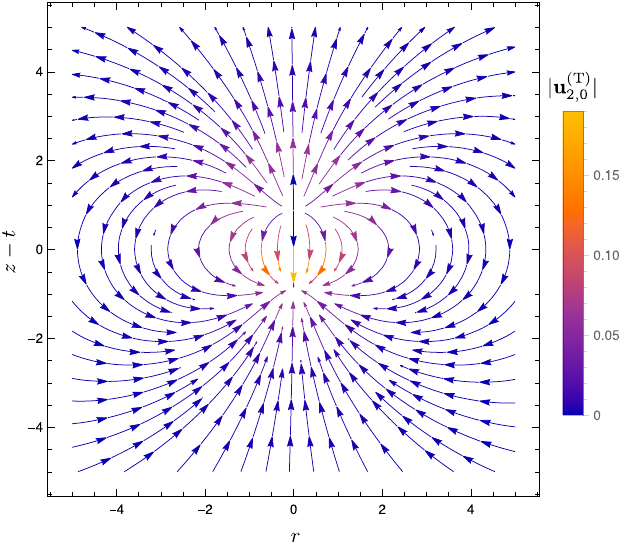}
		\caption{Streamlines of the axisymmetric flow $\mathbf{u}_{2,0}^\text{(T)}$ at order $\alpha^2$ due to a translating spherical heat spot during one scan. Colour indicates magnitude $\vert \mathbf{u}_{2,0}^\text{(T)} \vert $. 
		}
		\label{fig:plot_u200T_axisym_FLUCS_06022024}}
\end{figure}
As detailed above, we obtain the solution for the velocity field at order $\alpha^2$ as
\begin{align}
	\mathbf{u}_{2,0} \equiv A'(t)A(t) \mathbf{u}_{2,0}^\text{(S)} + A(t)^2 \mathbf{u}_{2,0}^\text{(T)},\label{eq:u20}
\end{align}
where the switching-on contribution is given by
\begin{align}
	\mathbf{u}_{2,0}^\text{(S)} 
	&= \left(\frac{\sqrt{2 \pi } \exp(- R^2/2) \erf(R/\sqrt{2})}{2 R^2}-\frac{\exp(-R^2)}{R}\right) \mathbf{e}_R,\label{eq:u20S}
\end{align}
and the translation contribution is given by
\begin{align}
	\mathbf{u}_{2,0}^\text{(T)} =
	\cos\theta \mathbf{e}_R  \Bigg(&
	\frac{\sqrt{\pi }  \erf(R)}{6 R^3}
	+\frac{\sqrt{2 \pi } \exp(- R^2/2) \erf(R/\sqrt{2})}{3 R} \nonumber\\
	&+\frac{2\sqrt{2 \pi }  \exp(- R^2/2) \erf(R/\sqrt{2})}{3R^3}
	 -  \exp(-R^2)
	-\frac{5  \exp(-R^2)}{3 R^2} \nonumber\\
	&
	+\frac{1}{6} \pi  [\erf(R/\sqrt{2})^2-1]
	\Bigg)\nonumber\\
	+ \sin\theta \mathbf{e}_\theta \Bigg(&
	\frac{\sqrt{\pi } \erf(R)}{12 R^3}
	-\frac{\sqrt{2 \pi } \exp(- R^2/2) \erf(R/\sqrt{2})}{3 R}
	+\frac{\sqrt{2 \pi } \exp(- R^2/2) \erf(R/\sqrt{2})}{3 R^3}\nonumber\\
	&-\frac{5 \exp(-R^2)}{6 R^2}
	-\frac{1}{6} \pi  [\erf(R/\sqrt{2})^2-1]
	\Bigg) ,\label{eq:u20T}
\end{align}
for $-t_0 \leq t \leq t_0$ (i.e.~during one scan).
We illustrate the radial flow associated with the switching-on of the heat spot $\mathbf{u}_{2,0}^\text{(S)}$ in Fig.~\ref{fig:plot_u200S_axisym_FLUCS_06022024}, and the dipolar flow associated with the translation of the heat spot $\mathbf{u}_{2,0}^\text{(T)}$ in Fig.~\ref{fig:plot_u200T_axisym_FLUCS_06022024}. 
These two flows are reminiscent of and may be viewed as reinforcing their counterparts at order $\alpha$; the flow at order $\alpha^2$ is the first correction to the flow at order $\alpha$ due to the fact that the density is lowered by heating below its reference value~$\rho_0$ for positive thermal expansion coefficient ($\alpha>0$), in the flux term $\nabla \cdot (\rho \mathbf{u})$ in mass conservation [Eq.~\eqref{eq:mass_consn_3D_axisym}].

The far-field behaviour of the flow $\mathbf{u}_{2,0}$ at order $\alpha^2$ is given by
\begin{align}
	\mathbf{u}_{2,0} \sim 
	\frac{A(t)^2\sqrt{\pi }}{12 R^3}
	(2\cos\theta \mathbf{e}_R  
	+ \sin\theta \mathbf{e}_\theta 
	).\label{eq:u20_far}
\end{align}
This is opposite in sign to the far field of order $\alpha\beta$, as for the parallel-plate setup~\cite{liao2023theoretical}.
However, deviating from the two-dimensional case~\cite{liao2023theoretical}, the numerical prefactor has a different magnitude compared with that at order $\alpha\beta$ for the axisymmetric setup without confinement; although much of the physics is shared between the different geometries, our calculations demonstrate explicitly how the details of the flow are specific to each geometry.

In dimensional terms, the flow during a scan at order $\alpha^2$ scales as $\alpha^2 \Delta T_0^2 U$; in contrast with order $\alpha\beta$, changing the sign of the thermal expansion coefficient does not affect the contribution at order $\alpha^2$.

Mathematically, for a general fluid, we will show that this purely thermal expansion-driven flow at order $\alpha^2$ gives rise to a contribution to the leading-order average velocity of tracers over a scan, typically in the opposite direction to the thermoviscous contribution at order $\alpha\beta$. 

\section{Net transport of tracers}\label{sec:transport}

We introduced in Sec.~\ref{sec:flow} our model for the thermoviscous and thermal expansion-driven flows induced by a translating spherical heat spot in three-dimensional, unbounded fluid, solving for the instantaneous flow during one scan of the heat spot, up to quadratic order in the thermal expansion coefficient $\alpha$ and thermal shear viscosity coefficient $\beta$.
In two-dimensional experiments, the relevant physical quantity observed is the net displacement of tracer beads due to repeated scanning of the laser, instead of the instantaneous fluid flow during one scan; furthermore, the theoretical average velocity of tracers from hydrodynamic modelling has been used quantitatively to design scan patterns to manipulate particles~\cite{liao2023theoretical,minopoli2023iso,erben2024opto,erben2025model}.
We therefore anticipate that future three-dimensional experiments, like their two-dimensional counterparts, will measure and exploit the time-averaged trajectories of tracers; consequently, predicting the net displacement of tracers due to our three-dimensional, unconfined fluid flow over the course of one scan is a key step towards applications in micromanipulation.
We thus now examine the kinematics of tracer particles in this flow and compute the net displacement of tracers due to a full scan of the heat spot from $z=-\ell$ to $z=\ell$.

\subsection{Trajectory of tracer}

We solve for the net displacement of a material point due to one full scan of the heat spot along the scan path from $z=-\ell$ to $z=\ell$.
Consider a material point that has initial position $\mathbf{X}_0 $ at time $t=-t_0$. 
Following the method in Ref.~\cite{liao2023theoretical}, we write its position vector relative to the origin at time $t$ as  $\mathbf{X}(t)$. 
In the absence of noise, this obeys an ordinary differential equation given by
\begin{align}
	\frac{d \mathbf{X}}{d t} &= \mathbf{u}(\mathbf{X}(t),t),
\end{align}
for $-t_0 \leq t \leq t_0$ (i.e.~during one scan).
The equivalent integral equation is given by
\begin{align}
	\mathbf{X}(t) - \mathbf{X}_0 = \int_{-t_0}^t \mathbf{u}(\mathbf{X}(\tilde{t}), \tilde{t}) \, d\tilde{t}.\label{eq:X_integral_eq}
\end{align}

\subsection{Perturbation expansion}

As in Ref.~\cite{liao2023theoretical}, we pose a perturbation expansion for the displacement vector $\Delta \mathbf{X}(t)\equiv \mathbf{X}(t) - \mathbf{X}_0$ as
\begin{align}
	\Delta \mathbf{X}(t) 
	=   \alpha \Delta\mathbf{X}_{1,0}(t) + \alpha^2 \Delta\mathbf{X}_{2,0}(t) + \alpha\beta \Delta\mathbf{X}_{1,1}(t) + \text{h.o.t.},\label{eq:displacement_pertn_exp}
\end{align}
where $\Delta\mathbf{X}_{m,n}(t) $ is the order-$\alpha^m\beta^n$ displacement  of the material point at time $t$ from the position $\mathbf{X}_0$ at $t=-t_0$.
Then expanding Eq.~\eqref{eq:X_integral_eq} yields
\begin{align}
	&\alpha \Delta\mathbf{X}_{1,0}(t) + \alpha^2 \Delta\mathbf{X}_{2,0}(t) + \alpha\beta \Delta\mathbf{X}_{1,1}(t) +\text{h.o.t.} \nonumber\\
	=&  \alpha\int_{-t_0}^t \mathbf{u}_{1,0}(\mathbf{X}_0,\tilde{t}) \, d\tilde{t} 
	+ \alpha^2\int_{-t_0}^t [\mathbf{u}_{2,0}(\mathbf{X}_0,\tilde{t}) + \Delta \mathbf{X}_{1,0}(\tilde{t}) \cdot \nabla \mathbf{u}_{1,0}(\mathbf{X}_0,\tilde{t})] \, d\tilde{t}\nonumber\\
	&+  \alpha\beta \int_{-t_0}^t \mathbf{u}_{1,1}(\mathbf{X}_0,\tilde{t}) \,d\tilde{t}.\label{eq:displacement_expanded}
\end{align}

\subsection{Zero net displacement at order $\alpha$}

We show here that the net displacement of a tracer after one full scan of the heat spot varies not linearly with the temperature change, but instead (at least) quadratically.
From the perturbation expansion in Eq.~\eqref{eq:displacement_expanded}, the displacement $\Delta\mathbf{X}_{1,0}(t)$ at order $\alpha$ (i.e.~leading order) of a material point at time $t$ is given by
\begin{align}
	\Delta \mathbf{X}_{1,0}(t) &= \int_{-t_0}^t \mathbf{u}_{1,0}(\mathbf{X}_0,\tilde{t}) \, d\tilde{t} .\label{eq:X10_integral}
\end{align}
Using Eq.~\eqref{eq:u100_general_decomposn} for $\mathbf{u}_{1,0}$ and the Fundamental Theorem of Calculus,  for a general heat spot, the expression for the order-$\alpha$ displacement in Eq.~\eqref{eq:X10_integral} becomes
\begin{align}
	\Delta \mathbf{X}_{1,0}(t) &= \int_{-t_0}^t \frac{\partial}{\partial \tilde{t}}  [A(\tilde{t}) \mathbf{u}_{1,0}^\text{(S)}(\mathbf{X}_0,\tilde{t})] \, d\tilde{t}\nonumber\\
	&= A(t)\mathbf{u}_{1,0}^\text{(S)}(\mathbf{X}_0,t) - A(-t_0)\mathbf{u}_{1,0}^\text{(S)}(\mathbf{X}_0,-t_0).
\end{align}
For a scan path of finite length, the heat-spot amplitude is zero at the ends of the scan path, so this simplifies to 
\begin{align}
	\Delta \mathbf{X}_{1,0}(t) &= A(t)\mathbf{u}_{1,0}^\text{(S)}(\mathbf{X}_0,t).\label{eq:X10_general}
\end{align}
Furthermore, the net displacement $\Delta \mathbf{X}_{1,0}(t_0)$ (due to a full scan) at order $\alpha$, of any material point, is given by
\begin{align}
	\Delta \mathbf{X}_{1,0}(t_0) = \mathbf{0}.
\end{align}
Thus, importantly, the leading-order net displacement of a tracer occurs not at linear order, but instead at quadratic order in the dimensionless parameters $\alpha$ and $\beta$; net transport is hence quadratic in the temperature perturbation.

\subsection{Net displacement at order $\alpha\beta$ and order $\alpha^2$}

Since we showed above that the net displacement at order $\alpha$ is precisely zero, the perturbation expansion for the net displacement of the material point [Eq.~\eqref{eq:displacement_pertn_exp}] can be rewritten as
\begin{align}
	\Delta \mathbf{X}(t_0) 
	=   \alpha^2 \Delta\mathbf{X}_{2,0}(t_0) + \alpha\beta \Delta\mathbf{X}_{1,1}(t_0) + \text{h.o.t.}\label{eq:net_displ_expansion}
\end{align}
Here, by Eq.~\eqref{eq:displacement_expanded}, the thermoviscous net displacement $\Delta\mathbf{X}_{1,1}(t_0)$ at order $\alpha\beta$ of a material point with initial position $\mathbf{X}_0$ is given by
\begin{align}
	\Delta\mathbf{X}_{1,1}(t_0) = \int_{-t_0}^{t_0} \mathbf{u}_{1,1}(\mathbf{X}_0,t) \,dt,\label{eq:net_displ_alphabeta}
\end{align}
and, similarly, the purely thermal expansion-driven net displacement~$\Delta\mathbf{X}_{2,0}(t_0)$ at order $\alpha^2$ is given by
\begin{align}
	\Delta\mathbf{X}_{2,0}(t_0) = \int_{-t_0}^{t_0} [\mathbf{u}_{2,0}(\mathbf{X}_0,t) + \Delta \mathbf{X}_{1,0}(t) \cdot \nabla \mathbf{u}_{1,0}(\mathbf{X}_0,t)] \, dt.\label{eq:net_displ_alphasq_integral}
\end{align}

\begin{figure}[t]
	\centering
	{\includegraphics[width=0.8\textwidth]{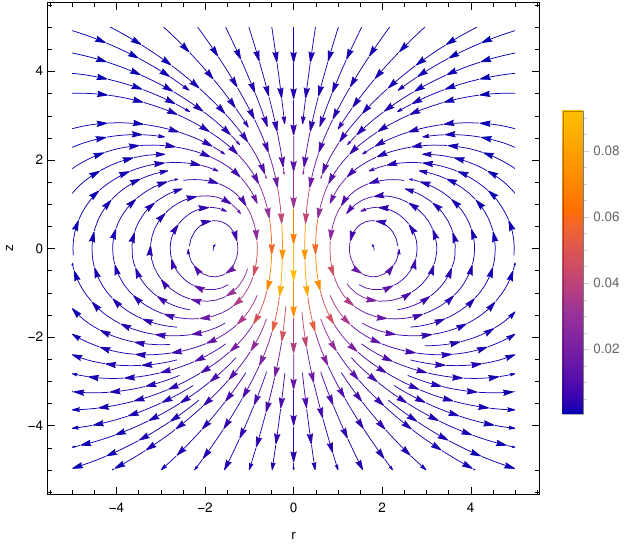}
		\caption{Net displacement $\Delta\mathbf{X}_{1,1}(t_0)$ of tracers at order $\alpha\beta$ due to one full scan of a spherical heat spot, with sinusoidal amplitude function [Eq.~\eqref{eq:A_sinusoidal}], and scan path from $z=-1.375$ to $z=1.375$ along the $z$ axis, with dimensionless scan-path length $2t_0= 2.75$. Colour indicates the magnitude~$\vert \Delta\mathbf{X}_{1,1}(t_0) \vert$.}
		\label{fig:plot_u110_transport_axisym_FLUCS_02022024}}
\end{figure}
\begin{figure}[t]
	\centering
	{\includegraphics[width=0.8\textwidth]{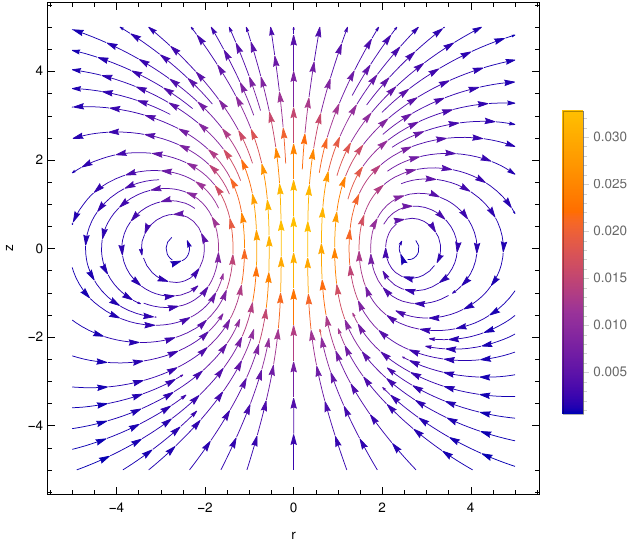}
		\caption{Net displacement $\Delta\mathbf{X}_{2,0}(t_0)$ of tracers at order $\alpha^2$ due to one scan of a spherical heat spot from $z=-1.375$ to $z=1.375$. Colour shows the magnitude $\vert \Delta\mathbf{X}_{2,0}(t_0) \vert$.}
		\label{fig:plot_u200_transport_axisym_FLUCS_06022024}}
\end{figure}

We plot the net displacement of tracers at order $\alpha\beta$ [Eq.~\eqref{eq:net_displ_alphabeta}] in Fig.~\ref{fig:plot_u110_transport_axisym_FLUCS_02022024} and at order $\alpha^2$ [Eq.~\eqref{eq:net_displ_alphasq_integral}] in Fig.~\ref{fig:plot_u200_transport_axisym_FLUCS_06022024}.
Here, to facilitate direct comparison with the results of Ref.~\cite{liao2023theoretical}, we choose a sinusoidal heat-spot amplitude function, given by
\begin{equation}
	A(t) = \cos^2 \left ( \frac{\pi t}{2 t_0} \right ),\label{eq:A_sinusoidal}
\end{equation}
for $-t_0 \leq t \leq t_0$, and we set the dimensionless scan-path length to be $2t_0= 2.75$, so that the scan path is from $z=-1.375$ to $z=1.375$ along the $z$ axis.
We note that the vertical axis label for these plots showing net transport of tracers due to a full scan is $z$, instead of $z-t$ used for the plots of instantaneous flow in Sec.~\ref{sec:flow}; that is, the net transport is shown here in the laboratory frame, as opposed to the co-moving frame. 

\subsubsection{Net transport near the scan path}

We now turn our attention to physical interpretation of our results on net transport [Eqs.~\eqref{eq:net_displ_alphabeta} and~\eqref{eq:net_displ_alphasq_integral}], illustrated in Figs.~\ref{fig:plot_u110_transport_axisym_FLUCS_02022024} and~\ref{fig:plot_u200_transport_axisym_FLUCS_06022024}, beginning with the behaviour close to the scan path.
Near the scan path, the upward translation of a heat spot thus produces two contributions to net transport of tracers, which typically have opposite directions, and scale dimensionally with~$\alpha\beta\Delta T_0^2$ and~$\alpha^2 \Delta T_0^2$.
First, assuming positive thermal expansion coefficient $\alpha$ and thermal shear viscosity coefficient $\beta$, the net transport at order $\alpha\beta$ near the scan path is in the opposite direction to heat-spot translation (Fig.~\ref{fig:plot_u110_transport_axisym_FLUCS_02022024}); this is an extension of the classic net thermoviscous flow of fluid confined between parallel plates and arises from the combination of thermal expansion and temperature-dependent shear viscosity~\cite{weinert2008optically,erben2021feedback,liao2023theoretical,weinert2008microscale}. 
The physical mechanism for this thermoviscous net flow is inherited via the time-averaging in Eq.~\eqref{eq:net_displ_alphabeta} from the instantaneous flow at order $\alpha\beta$, explained in Sec.~\ref{sec:u11_physical_interpret}.
A negative thermal expansion coefficient $\alpha$ (e.g.~for water below $4~\si{\degreeCelsius}$) would reverse the direction of the contribution at order $\alpha\beta$ to net displacement of tracers, by changing the sign of $\alpha\beta$ in the perturbation expansion in Eq.~\eqref{eq:net_displ_expansion}.

Secondly, thermal expansion associated with the scanning heat spot can, by itself, give rise to nonzero net transport at order $\alpha^2$, via a fundamentally different physical mechanism.
This is independent of thermal shear viscosity changes and is inherited from the flow at order $\alpha^2$ during one scan (Sec.~\ref{sec:alpha_sq}). 
In principle, achieving net transport therefore does not require temperature-dependent viscosity, only thermal expansion, though in practice, its significance would vary based on the material parameters of the liquid used.
This contribution is characterised by net transport always in the same direction as the heat-spot translation (upwards), whether the thermal expansion coefficient $\alpha$ is positive or negative.
For aqueous glycerol, we expect the thermoviscous net transport at order $\alpha\beta$ to dominate over the purely thermal expansion-driven contribution at order $\alpha^2$. 
However, if a liquid has sufficiently large thermal expansion coefficient relative to the thermal shear viscosity coefficient, then the net transport at order $\alpha^2$ may dominate instead, with opposite direction. 
In all cases, there is no contribution from the bulk viscosity.

\subsubsection{Far-field behaviour}

We finally examine the far-field behaviour of the thermoviscous and thermal expansion-driven net transport. 
In the far field, the net displacement at order $\alpha\beta$ is given by
\begin{align}
	\Delta \mathbf{X}_{1,1}(t_0) \sim -\frac{ \sqrt{\pi}}{20} \int_{-t_0}^{t_0} A(t)^2 \, dt \left . \left (\frac{ 1}{R^3} ( 2  \cos\theta \mathbf{e}_R+  \sin\theta \mathbf{e}_\theta) \right ) \right \vert_{t=-t_0},\label{eq:X11_far}
\end{align}
which is inherited from the source dipole in the far field of the flow $\mathbf{u}_{1,1}$ via Eqs.~\eqref{eq:net_displ_alphabeta},~\eqref{eq:u11} and~\eqref{eq:u11_far}, with physical interpretation given in Sec.~\ref{sec:u11_physical_interpret}.
We note that evaluating at $t=-t_0$ corresponds to spherical polar coordinates with origin at $(r=0,z=-t_0)$ (in terms of cylindrical coordinates), with the tracer at its initial position. 
Similarly, the far-field net displacement at order $\alpha^2$ is given by
\begin{align}
	\Delta \mathbf{X}_{2,0}(t_0) \sim \frac{\sqrt{\pi }}{12} \int_{-t_0}^{t_0} A(t)^2 \, dt
	\left . \left (\frac{1}{R^3}
	(2\cos\theta \mathbf{e}_R  
	+ \sin\theta \mathbf{e}_\theta 
	) \right )\right \vert_{t=-t_0}.\label{eq:X20_far}
\end{align}
Here, we have considered the contribution to the far-field spatial decay of the net displacement in Eq.~\eqref{eq:net_displ_alphasq_integral} from Eqs.~\eqref{eq:u20}--\eqref{eq:u20T} for the flow at order $\alpha^2$, Eq.~\eqref{eq:X10_general} for the displacement of a tracer at order $\alpha$, and Eqs.~\eqref{eq:u100_general_decomposn},~\eqref{eq:u10S} and~\eqref{eq:u10T} for the flow at order~$\alpha$.
The result for the far-field net displacement at order $\alpha^2$ in Eq.~\eqref{eq:X20_far} originates from the far-field source dipole in the translation contribution $\mathbf{u}_{2,0}^\text{(T)}$ to flow during a scan at order $\alpha^2$ [Eq.~\eqref{eq:u20_far}], explained physically in Sec.~\ref{sec:u20_result}.

Combining Eqs.~\eqref{eq:X11_far} and~\eqref{eq:X20_far}, we therefore obtain the dimensional far-field net displacement of a tracer as
\begin{align}
	\Delta\mathbf{X}(t_0) 
	\sim \sqrt{\pi}\left (-\frac{ 1}{20}\alpha\beta+\frac{1}{12}\alpha^2\right )\Delta T_0^2 U \int_{-t_0}^{t_0} A(t)^2 \, dt 	\left . \left (\frac{a^3}{R^3}
	(2\cos\theta \mathbf{e}_R  
	+ \sin\theta \mathbf{e}_\theta 
	) \right )\right \vert_{t=-t_0}.	
\end{align}
Hence, the far-field average velocity, over one scan, of the material point with initial position $\mathbf{X}_0$ is given by
\begin{align}
	\frac{\Delta\mathbf{X}(t_0)}{2t_0} 
	\sim \sqrt{\pi}\left (-\frac{ 1}{40}\alpha\beta+\frac{1}{24}\alpha^2\right ) \frac{\Delta T_0^2 U}{t_0} \int_{-t_0}^{t_0} A(t)^2 \, dt 	\left . \left (\frac{a^3}{R^3}
	(2\cos\theta \mathbf{e}_R  
	+ \sin\theta \mathbf{e}_\theta 
	) \right )\right \vert_{t=-t_0}.	\label{eq:avg_vel_far-field_quadratic}
\end{align}
This is a hydrodynamic source dipole in three dimensions. 
Its strength scales quadratically with the peak temperature change~$\Delta T_0$ and linearly with the speed of heat-spot translation $U$ (and hence linearly with frequency of scanning), while the direction depends on the thermal expansion coefficient~$\alpha$ and thermal shear viscosity coefficient~$\beta$ purely through the factor of $\left (-\frac{ 1}{40}\alpha\beta+\frac{1}{24}\alpha^2\right )$.
These two terms reflect and quantify the two different physical effects driving net transport: the interplay between thermal expansion and thermal shear viscosity changes at order $\alpha\beta$, and thermal expansion by itself at order $\alpha^2$.
If the thermal expansion coefficient $\alpha$ and thermal shear viscosity coefficient $\beta$ are both positive, as is the case for many liquids, then the two effects compete; the larger prefactor for the order-$\alpha^2$ term could potentially compensate for a value of $\alpha$ smaller than $\beta$. 
However, for a negative thermal expansion coefficient $\alpha$, the two terms in the prefactor are instead of the same sign, thus reinforcing each other.

\section{Discussion}\label{sec:discussion}

In this article, we considered thermoviscous and thermal expansion-driven fluid flow in three-dimensional, unbounded fluid: a new geometry compared with previous work on viscous fluid confined between parallel plates~\cite{liao2023theoretical}. 
We first examined heat transport. By solving numerically for the temperature field induced by a scanning heat source in the limit relevant to experiments~\cite{mittasch2018non,minopoli2023iso}, we motivated a simplified model of the temperature perturbation to act as an input to our flow model. 
We then derived analytically the fluid flow and net transport due to the scanning heat spot, to quadratic order in the thermal expansion coefficient $\alpha$ and thermal shear viscosity coefficient $\beta$.
Our model included bulk viscosity, a key new physical ingredient present because of the compressible nature of the flow. 
Bulk viscosity did not play a part in the parallel-plate setup due to the geometry, as shown using a scaling argument~\cite{weinert2008microscale,liao2023theoretical}.
Here, for the three-dimensional, unconfined fluid, we treated the bulk viscosity as a general function of space and time, without specifying its temperature dependence. 
Through our analysis, we found that even though the bulk viscosity impacts the pressure field, it does not influence the fluid velocity.
We obtained the same physics and qualitative results as the parallel-plate setup.
Specifically, the leading-order instantaneous flow during one scan is driven by thermal expansion via the spatio-temporally varying temperature field, occurring at order $\alpha$ and linear in the peak temperature change. 
In contrast with this, the leading-order net transport occurs at both order~$\alpha\beta$ and order~$\alpha^2$, quadratic in the temperature change.
The two typically competing contributions at these two orders originate from two fundamentally different physical mechanisms, with the order-$\alpha\beta$ thermoviscous term generated by the interplay between thermal expansion and thermal shear viscosity changes, while the order-$\alpha^2$ effect arises purely from thermal expansion.
The far-field average velocity of tracers is given by a hydrodynamic source dipole, but now in three dimensions. 

We comment on the validity of our spherical approximation for the heat spot. 
In existing microfluidic experiments, the scan path can have length on the order of $50~\si{\micro\metre}$, but there can be a relatively large length scale of heat absorption by the fluid, perpendicular to the scan path, of around $300~\si{\micro\metre}$. 
However, for a scan path long enough that heat spot is small in comparison, the geometry would be approximately axisymmetric; our model could  therefore serve as a first approximation in this situation. 
Furthermore, methods of achieving thermoviscous transport in three dimensions that have been suggested include highly focused heating of the fluid~\cite{weinert2008optically}.

Our work provides a first model of fully three-dimensional thermoviscous and thermal expansion-driven net flows, which we anticipate will be key to explaining experimental data (Moritz Kreysing, personal communication).
The unconfined geometry of our model may help with understanding experiments where boundaries are far from the heat spot~\cite{mittasch2018non} or where the lubrication approximation (employed in theory for the parallel-plate setup~\cite{liao2023theoretical,erben2021feedback}) is no longer valid.
Our theory will also allow us to analyse the effect of boundaries and confinement on three-dimensional thermoviscous and thermal expansion flows and transport in future modelling work.
The quantitative theoretical results, in combination with scan-path selection via feedback algorithms~\cite{stoev2021highly,erben2021feedback,erben2024opto} or global optimisation~\cite{erben2025model}, could contribute to the design of new experiments that apply net thermoviscous and thermal expansion flows to trap or manipulate particles in three dimensions.

\backmatter

\bmhead{Acknowledgements}

We thank Moritz Kreysing for helpful discussions and feedback.
We gratefully acknowledge funding from the Engineering and Physical Sciences Research Council (studentship to W.L.) and Trinity College, Cambridge (Rouse Ball and Eddington Research Funds travel grant to W.L.). 

\bigskip

\bibliography{cytoplasmic_streaming_bibliography_22072024_axisym_FLUCS_20112024.bib}

\end{document}